\documentclass[10pt]{article}
\usepackage{booktabs}
\usepackage{setspace}
\usepackage{graphicx}%
\usepackage{multirow}%
\usepackage{amsmath,amssymb,amsfonts}%
\usepackage{amsthm}%
\usepackage{mathrsfs}%
\usepackage[title]{appendix}%
\usepackage{xcolor}%
\usepackage{textcomp}%
\usepackage{manyfoot}%
\usepackage{booktabs}%
\usepackage{algorithm}%
\usepackage{algorithmicx}%
\usepackage{algpseudocode}%
\usepackage{listings}%
\usepackage{caption}
\usepackage{subcaption}
\usepackage{natbib}
\usepackage[hidelinks,colorlinks=true,linkcolor=blue]{hyperref}
\hypersetup{citecolor=blue, urlcolor=blue}
\usepackage{geometry}
\usepackage{bm}

\theoremstyle{thmstyleone}%
\theoremstyle{thmstyletwo}%

\theoremstyle{thmstylethree}%

\usepackage{easy-todo}

\makeatletter
\ifcase \@ptsize \relax
  \newcommand{\miniscule}{\@setfontsize\miniscule{4}{5}}
\or
  \newcommand{\miniscule}{\@setfontsize\miniscule{5}{6}}
\or
  \newcommand{\miniscule}{\@setfontsize\miniscule{5}{6}}
\fi
\makeatother

\usepackage{lmodern}

\usepackage{lipsum}

\definecolor{codegreen}{rgb}{0,0.6,0}
\definecolor{codegray}{rgb}{0.5,0.5,0.5}
\definecolor{codepurple}{rgb}{0.58,0,0.82}
\definecolor{backcolour}{rgb}{0.95,0.95,0.92}

\lstdefinestyle{DOS}
{
    backgroundcolor=\color{black},
    basicstyle=\scriptsize\color{white}\ttfamily
}

\lstdefinestyle{mystyle}{
    backgroundcolor=\color{backcolour},   
    commentstyle=\color{codegreen},
    keywordstyle=\color{magenta},
    numberstyle=\tiny\color{codegray},
    stringstyle=\color{codepurple},
    basicstyle=\ttfamily\footnotesize,
    breakatwhitespace=false,         
    breaklines=true,                 
    captionpos=b,                    
    keepspaces=true,                 
    numbers=left,                    
    numbersep=5pt,                  
    showspaces=false,                
    showstringspaces=false,
    showtabs=false,                  
    tabsize=2
}

\usepackage{caption}
\usepackage{subcaption}
\usepackage{comment}
\title{Performance of an open-source image-based history matching framework for CO$_2$ storage}

\author{D. Landa-Marb\'an$^{1}$ \and
  T.H. Sandve$^{1}$ \and
  J.W. Both$^{2}$ \and
  J.M. Nordbotten$^{1,2}$ \and
  S.E. Gasda$^{1,3}$}
  
\date{}
\begin{document}
\pagenumbering{arabic}
\maketitle
\onehalfspace
\noindent ${}^1$ Division of Energy and Technology, NORCE Research AS, Nyg{\aa}rdsgaten 112, Bergen 5008, Norway.\\[5pt]
${}^2$ Center for Modeling of Coupled Subsurface Dynamics, Department of Mathematics, University of Bergen, All\'egaten 41, Bergen 5007, Norway.\\[5pt]
${}^3$ Department of Physics and Technology, University of Bergen, All\'egaten 41, Bergen 5007, Norway.\\[5pt]
Corresponding author: David Landa-Marb\'an (E-mail address: dmar@norceresearch.no).
\vspace{1.2cm}

\begin{abstract}
\noindent We present a history matching (HM) workflow applied to the International FluidFlower benchmark study dataset, which features high-resolution images of CO$_2$ storage in a meter-scale, geologically complex reservoir. The dataset provides dense spatial and temporal observations of fluid displacement, offering a rare opportunity to validate and enhance HM techniques for geological carbon storage (GCS). The combination of detailed experimental data and direct visual observation of flow behavior at this scale is novel and valuable. This study explores the potential and limitations of using experimental data to calibrate standard models for GCS simulation. By leveraging high-resolution images and resulting interpretations of fluid phase distributions, we adjust uncertain parameters and reduce the mismatch between simulation results and observed data. Simulations are performed using the open-source OPM Flow simulator, while the open-source Everest decision-making tool is employed to conduct the HM. After the HM process, the final simulation results show good agreement with the experimental CO$_2$ storage data. This suggests that the system can be effectively described using standard flow equations, conventional saturation functions, and typical PVT properties for CO$_2$-brine mixtures. Our results demonstrate that the Wasserstein distance is a particularly effective metric for matching multi-phase, multi-component flow data. The entire workflow is implemented in a Python package named \texttt{pofff} (Python OPM Flow FluidFlower), which organizes all functionality through a single input file. This design ensures reproducibility and facilitates future extensions of the study.
\end{abstract}


\section*{Highlights}
\begin{itemize}
\item The performance of OPM Flow enables rapid history matching studies on the FluidFlower (a single run completing in ca. 2 minutes)
\item Calibrated parameters are sensitive to grid type and resolution, i.e., parameter validity is restricted to the specific input grid used
\item The workflow for conducting history matching studies is provided in a user-friendly Python package named \texttt{pofff}
\end{itemize}

\section{Introduction}\label{sec1}
Geological carbon storage (GCS) is considered a crucial strategy for reducing emissions and achieving net-zero targets, supported by more than 30 years of pilot and commercial operations worldwide (\cite{CCSinstitute}). In recent years, interest in GCS has grown significantly, with several major projects either underway or in development. Notable examples include Quest in Canada, Porthos in the Netherlands, and Northern Lights in Norway (\cite{Furrue2019}). The collective experience from these diverse projects, spanning a range of geological and technical settings, has been complemented by multiple decades of research and insights from controlled laboratory experiments and modeling studies. Together, these efforts have contributed to a robust and evolving knowledge base on GCS processes. This understanding has been incorporated into both commercial and research-oriented simulation tools designed to model CO$_{2}$ migration and trapping within storage formations (\cite{RASMUSSEN2021159,mrst,KOCH2021423,open-DARTS}). These models are used in conjunction with detailed geological characterization and monitoring data throughout all stages of project development and operation. Applications include estimating storage capacity and injectivity, optimizing injection strategies, and supporting risk assessment and management.

Predicting GCS performance in realistic geological settings remains a significant challenge. This difficulty arises primarily from two sources of uncertainty. First, reservoir models are typically constructed using sparse well data and relatively low-resolution seismic surveys, which introduce substantial uncertainty into model parameters. Second, simulation tools often rely on simplified physical approximations to maintain computational efficiency, leading to modeling errors. As an example, a common simplification is to neglect capillary pressure at the field scale (\cite{NI2025105257}). Measured data play a crucial role in reducing both types of uncertainty and enhancing predictive accuracy. History matching (HM) algorithms, reducing the mismatch between collected data and simulation results, offer a systematic approach to calibrating model parameters by treating the reservoir simulator as a black box. In addition to parameter estimation, the resulting optimized simulation results and minimal mismatch can reveal missing or poorly represented physical processes within the underlying mathematical model. 

Several field studies have shown that matching field data often requires a combination of parameter calibration and adjustments to the mathematical model itself. A prominent example is the Sleipner benchmark dataset (\cite{sleipner}), where time-lapse seismic data have been used to calibrate formation properties, reservoir structure, geothermal gradients, and fluid composition, with varying levels of success (\cite{ieaghg2021}). These studies highlight the dominant role of gravity in the Utsira formation and suggest that HM alone is insufficient to address the limitations of standard simulators in capturing the strong gravity-driven segregation and migration patterns observed in the seismic data. Other notable examples include the In Salah project (\cite{RINGROSE20136226}), where the assumption of linearly elastic deformation failed to reproduce observed surface uplift suggesting the use of nonlinear stress-strain relationships (\cite{RINALDI20173247}), and the Tub\r{a}en injection at Sn{\o}hvit, where matching downhole pressure data required incorporating additional geological heterogeneity, such as sub-seismic faults (\cite{HANSEN20133565}), and physical processes like salt precipitation (\cite{GRUDE2014178}).

Solubility trapping is a key mechanism in GCS, where the dissolution of CO$_2$ into formation brine is enhanced by density-driven convection. As CO$_2$ dissolves, the brine becomes denser, leading to gravitational instabilities that cause the CO$_2$-rich brine to sink in finger-like patterns (\cite{ennis2003role}). This convective mixing process has been extensively studied in controlled laboratory settings, including Hele-Shaw cell experiments conducted under relevant temperature and pressure conditions (\cite{faisal2015quantitative, amarasinghe2020effects}), as well as in intermediate-scale porous media experiments (\cite{AGARTAN2020102888}) and numerical simulations (\cite{neufeld2010convective, elenius2013convective}). These studies have primarily focused on the dissolution process in both homogeneous and heterogeneous media, often in isolation from other migration and trapping mechanisms. While experimental results suggest that convective mixing should occur at the field scale, direct observation of this phenomenon in subsurface reservoirs remains elusive. At the Sleipner site, gravimetric data have been used to estimate an upper bound on dissolution rates due to convective mixing. Subsequent simulation studies have supported the plausibility of these estimates by comparing modeled results with seismic observations (\cite{Mykkeltvedt2012}).

Experimental studies have played a key role in constraining model assumptions and parameters under idealized conditions. However, verifying and calibrating simulation models that account for convective mixing alongside other CO$_2$ storage processes in realistic geological settings remains a significant challenge. This difficulty stems from the limited availability of data that capture convective mixing in conjunction with CO$_2$ migration influenced by factors such as injection dynamics, formation and petrophysical properties, residual trapping, permeability and capillary barriers, and fault-related fluid flow. Recent advances have addressed this gap through the FluidFlower experimental system, a meter-scale CO$_2$ injection rig designed to replicate dominating physical processes during geological CO$_2$ storage~(\cite{ferno2024}). At this scale, model parameters and injection conditions can be precisely measured and controlled. The facility enables direct observation of CO$_2$ migration and dissolution using high-resolution visual imaging, resulting in a spatially dense, time-lapse dataset that captures the dynamic behavior of CO$_2$ in porous media. Accompanied with calibrated image analysis tools~(\cite{DarSIA}), quantitative datasets can be generated enabling comparison against simulation results. In this spirit, the double-blind, community-wide FluidFlower Validation Benchmark Study has been organized~(\cite{Nordbotten:Manual:2022}), comprising of five repeated multi-day experiments of CO$_2$ storage in a meter-scale, layered, heterogeneous sand geometry. Nine international modeling groups participated aiming at replicating the experiments without being able to review the experimental results before submitting their results. Finally, quantitative comparisons were drawn between the modeling and experimental results~(\cite{Flemisch2024}) and the accompanying image data, processing and data comparison scripts have been published.

The International FluidFlower benchmark study dataset~(\cite{eikehaug_2023_7510589}) offers a unique opportunity to investigate a key open question in GCS: how does convective mixing interact with other multi-phase flow processes, and can standard simulators accurately capture this interplay? Specifically, the study examines whether physical mechanisms that are currently missing from existing models need to be incorporated, or if improved predictability can be achieved through the careful calibration of model inputs using experimental data. To address this question, we apply a HM workflow to time-lapse images from the series of CO$_2$ injection experiments associated to the FluidFlower benchmark. On the one hand, this approach enables direct comparison between simulated and observed flow behavior, providing insight into the adequacy of current modeling frameworks. On the other hand, through the systematic approach and with full data visibility, this study directly complements the community-wide benchmark initiative~(\cite{Flemisch2024}), providing the possibility to provide the first "optimal" modeling fit of the dataset. Furthermore, this also clarifies whether parameter tuning is a key factor in aligning the model with the experimental data, especially in contrast to the double-blind benchmark participants. 

Regarding numerical simulations, reproducibility remains a significant challenge in research published in academic journals. For example, a recent study by \cite{Riehl2025} examined 11,879 simulation studies and 672 associated repositories in the field of transportation modeling, finding that fewer than 2 percent provided supplementary materials to support their findings -- a trend we must also assume conceptually applies to the field of reservoir modeling. This highlights a broader issue across scientific disciplines, including reservoir engineering. As a key contribution of this work, we introduce an open-source workflow named \texttt{pofff} (Python OPM Flow FluidFlower), which is hosted in a public repository~(\cite{pofff}). This tool adopts the workflow and methods from \texttt{pyopmspe11}, a Python framework using OPM Flow for the SPE11 benchmark project (\cite{Landa-Marbán2025}), aiming at multiphase flow simulations of FluidFlower experiments. The repository contains the complete model setup, simulation tools, and HM framework, enabling researchers to reproduce the results presented in this study. Moreover, the workflow is designed to be adaptable for future experiments conducted in the FluidFlower facility~(\cite{Eikehaug2024}) or similar experimental rigs. This resource aims to support ongoing research efforts and serve as a practical training tool for the broader scientific community.

The remainder of the paper is organized as follows. Section 2 outlines the computational workflow, including a description of the experimental dataset, the mathematical model used for flow simulations, and the algorithm employed for HM. Section 3 presents the HM results and compares them with previously published outcomes from the FluidFlower benchmark study. Section 4 offers concluding remarks and discusses directions for future research. The Appendix includes an example input configuration file for the \texttt{pofff} workflow, as well as illustrative model sensitivity results.

\section{Methodology}\label{sec2}
In this section we describe the followed steps from the design of the FluidFlower benchmark experiments to the development of the Python OPM Flow FluidFlower \texttt{pofff} simulation tool.

\subsection{Experimental FluidFlower data}\label{sec2.1}
This subsection summarizes the laboratory CO$_2$ storage experiments conducted as part of the FluidFlower benchmark. A detailed description of these, allowing to setup associated simulations, is provided in \cite{Nordbotten:Manual:2022}, and the experimental data is presented in~\cite{ferno2024}.

\begin{figure}[h!]
\centering
\includegraphics[width=\textwidth]{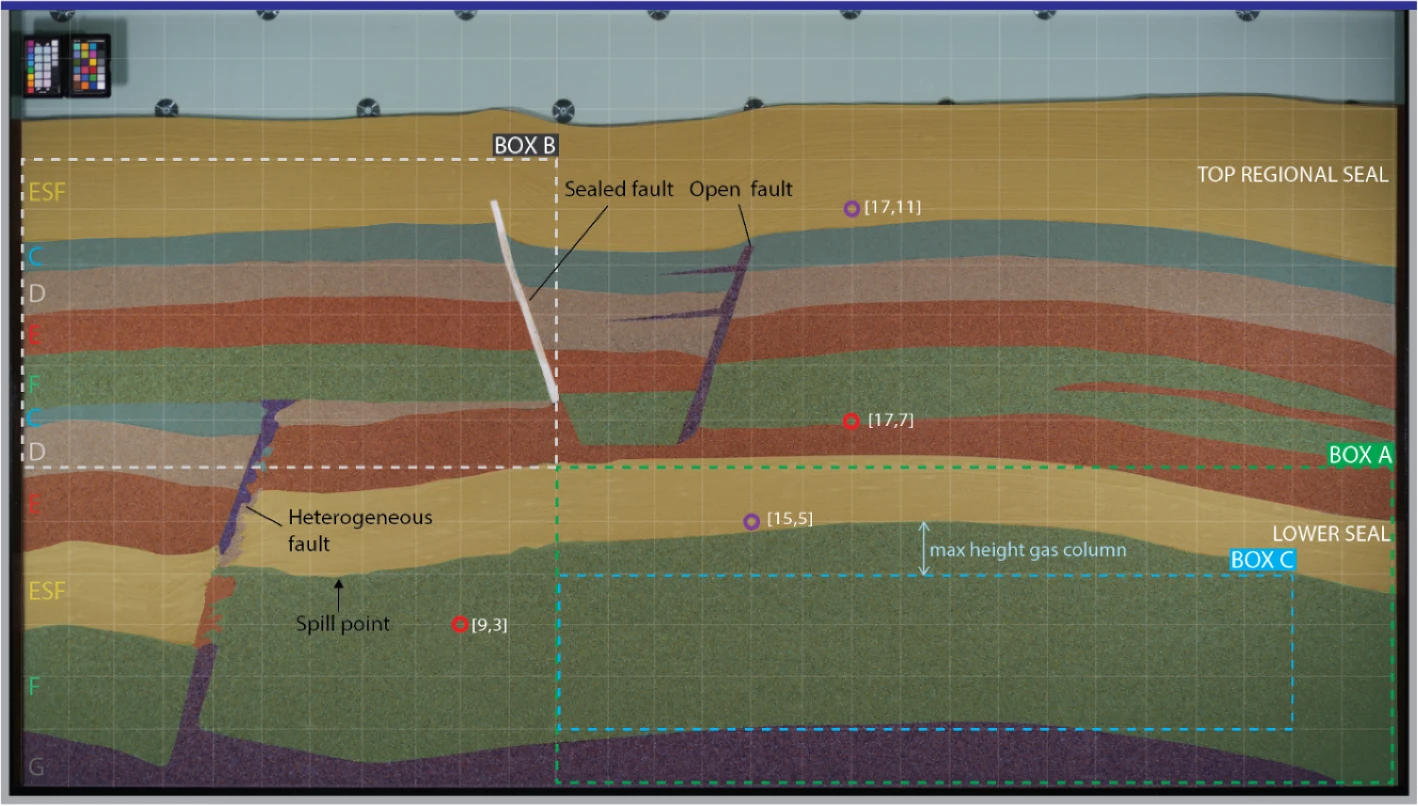}
\caption{The FluidFlower benchmark geometry. Figure courtesy of~\cite{ferno2024}}
\label{ff_rig}
\end{figure}

The FluidFlower is a meter-scale CO$_2$ injection rig featuring a complex geometric design and transparent glass walls, enabling direct observation and high-resolution data acquisition of CO$_2$ migration. The experimental setup of the FluidFlower is described in more detail by \cite{ferno2024}, \cite{Haugen2024}, and \cite{Eikehaug2024}. The geometry of the FluidFlower rig, used in this study, is illustrated in Figure~\ref{ff_rig}.

The experimental domain has a length of 2.8 meters and a height of 1.5 meters. The initial thickness (before running the CO$_2$ experiments) varies from 0.019 meters at the sides to a maximum of 0.028 meters at the center (see Figure 3 in \cite{Nordbotten:Manual:2022}). The injection ports, indicated by red circles in the figure, have a radius of 0.0009 meters. The density of the water used in the experiments is approximately 1002 kg/m$^3$. All experiments were conducted at ambient room temperature and pressure.

The experimental system consists of six distinct sand types and includes a silicon bar, represented by the white strip in box B of Figure~\ref{ff_rig}, which acts as a blocking fault. The entire domain is saturated with water containing a pH-sensitive dye, enabling visual tracking of CO$_2$ displacement. This setup allows time-lapse photographs to serve as measurement data for comparison with simulation results. The domain is divided into three regions, referred to as boxes A, B, and C in Figure~\ref{ff_rig}. These regions are used to compute spatially averaged quantities from the simulations, such as CO$_2$ concentration, which are discussed in the numerical results presented in Section \ref{sec3}.

\subsubsection{Sand properties}
In Figure~\ref{ff_rig}, the sand with the largest grain size (approximately 2.5 mm), which also has the highest absolute permeability, is represented by the dark purple color. This sand is primarily located along the bottom of the domain and in two diagonal regions that simulate high-permeability fault zones. In contrast, the finest sand (approximately 0.2 mm) is positioned at two distinct elevations along the length of the domain to represent caprock layers that act as vertical flow barriers. The measured physical properties for each sand type used in the simulations are summarized in Table~\ref{parameters}.

\begin{table}[h]
\centering
\caption{Measured sand properties (\cite{Nordbotten:Manual:2022})}
\label{parameters}
\begin{tabular}{@{}lrlrlllllr@{}}		
\toprule
Id & Sand& Grain size [mm] &$k$ [D]&$\phi$  [-] &$s_{r,w}$ [-]&$k^e_{r,g}$ [-]&  $s_{r,g}$ [-]  & $k^e_{r,w}$  & $p_e$ [Pa]\\
\midrule
1 & ESF	&0.20$\pm$0.11&44	 	&0.43&0.32&0.09& 0.14& 0.71&1471.5\\
2 & C	&0.66$\pm$0.09&473	&0.44&0.14&0.05& 0.10& 0.93&294.3\\
3 & D	&1.05$\pm$0.14&1110	&0.44&0.12&0.02& 0.08& 0.95&98.1\\
4 & E	&1.45$\pm$0.19&2005	&0.45&0.12&0.10& 0.06& 0.93&  -\\
5 & F	&1.77$\pm$0.31&4259 	&0.45&0.12&0.11& 0.13& 0.72&-\\
6 & G	&2.51$\pm$0.63&9580	&0.44&0.10&0.16& 0.06& 0.75&-\\
\bottomrule
\end{tabular}
\end{table}

\noindent In Table~\ref{parameters}, $k$ denotes the absolute permeability, while $\phi$ represents the porosity. The parameter $s_{r,w}$ corresponds to the residual water saturation, below which the water phase is immobile. Similarly, $s_{r,g}$ is the residual gas saturation, defining the threshold below which the gas phase is immobile. The terms $k^e_{r,g}$ and $k^e_{r,w}$ indicate the endpoint relative permeabilities of gas and water, respectively. Finally, $p_e$ refers to the capillary entry pressure.

As shown in Table~\ref{parameters}, entry pressure measurements are not available for Sands E, F, and G. This is due to limitations in the measurement technique. For reference, the entry pressure of 98.1 Pa reported for Sand D corresponds to a measured gas column height of only 0.01 meters. Given the sensitivity of fluid distribution and composition to entry pressure values, these parameters play a critical role in the simulation and must be estimated through the HM process.

\subsubsection{CO$_2$ experiments and data acquisition}
\begin{figure}[h!]
\centering
\begin{subfigure}[b]{0.32\textwidth}\includegraphics[width=\textwidth]{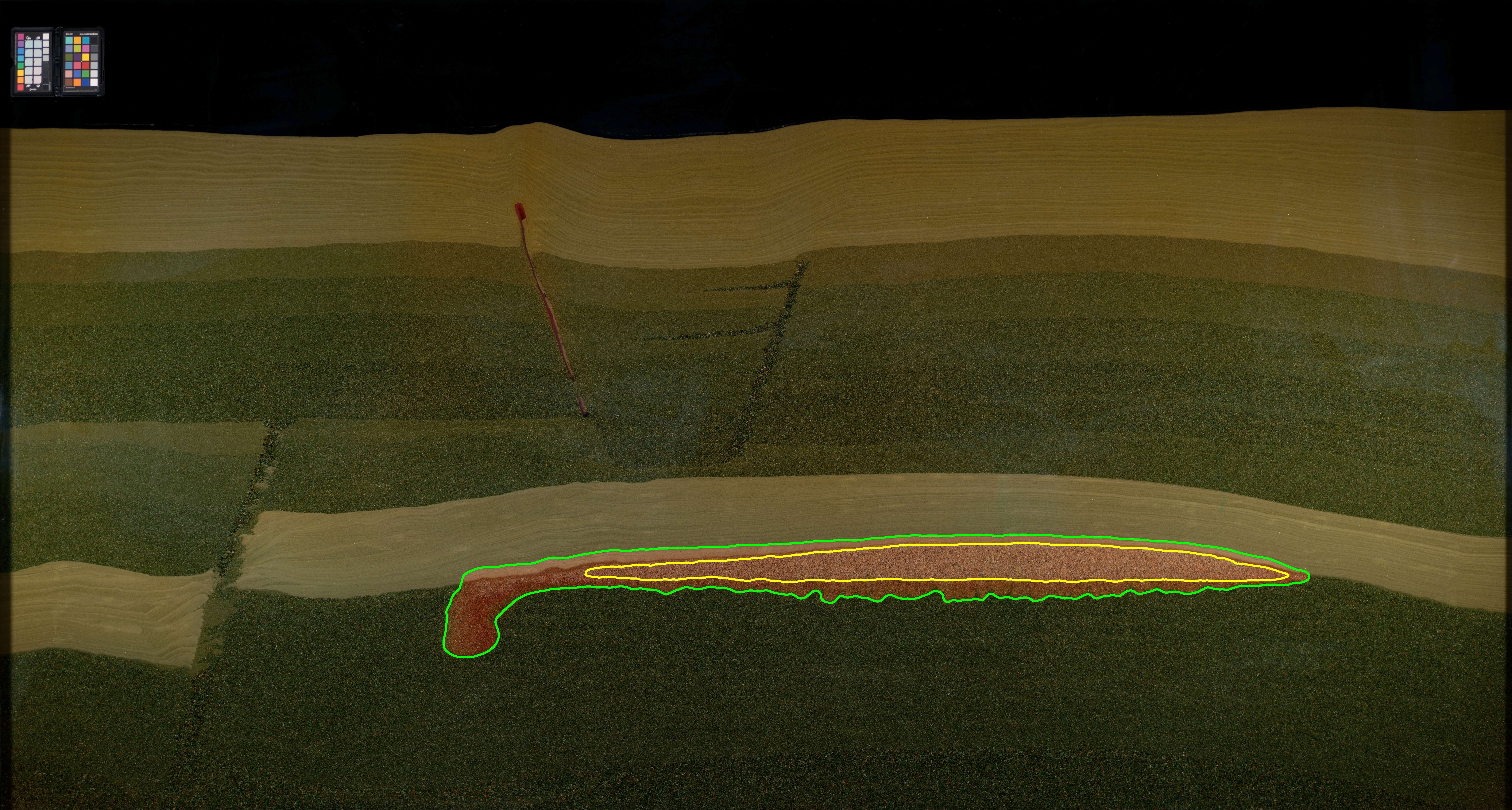}
\caption{2 hrs 15 min}
\label{co2experimenta}
\end{subfigure}
\begin{subfigure}[b]{0.32\textwidth}\includegraphics[width=\textwidth]{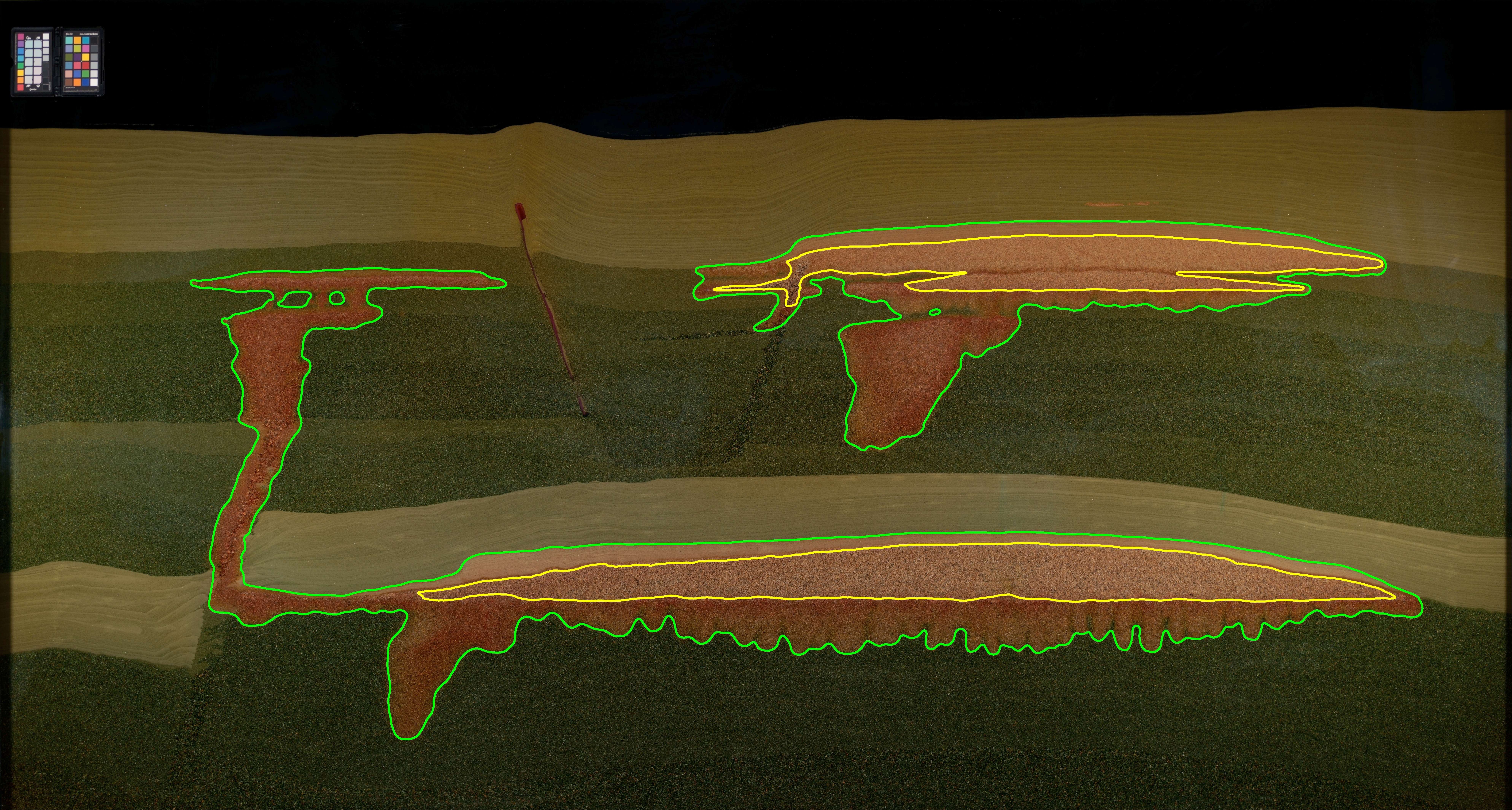}
\caption{5 hrs 5 min}
\label{co2experimentb}
\end{subfigure}
\begin{subfigure}[b]{0.32\textwidth}\includegraphics[width=\textwidth]{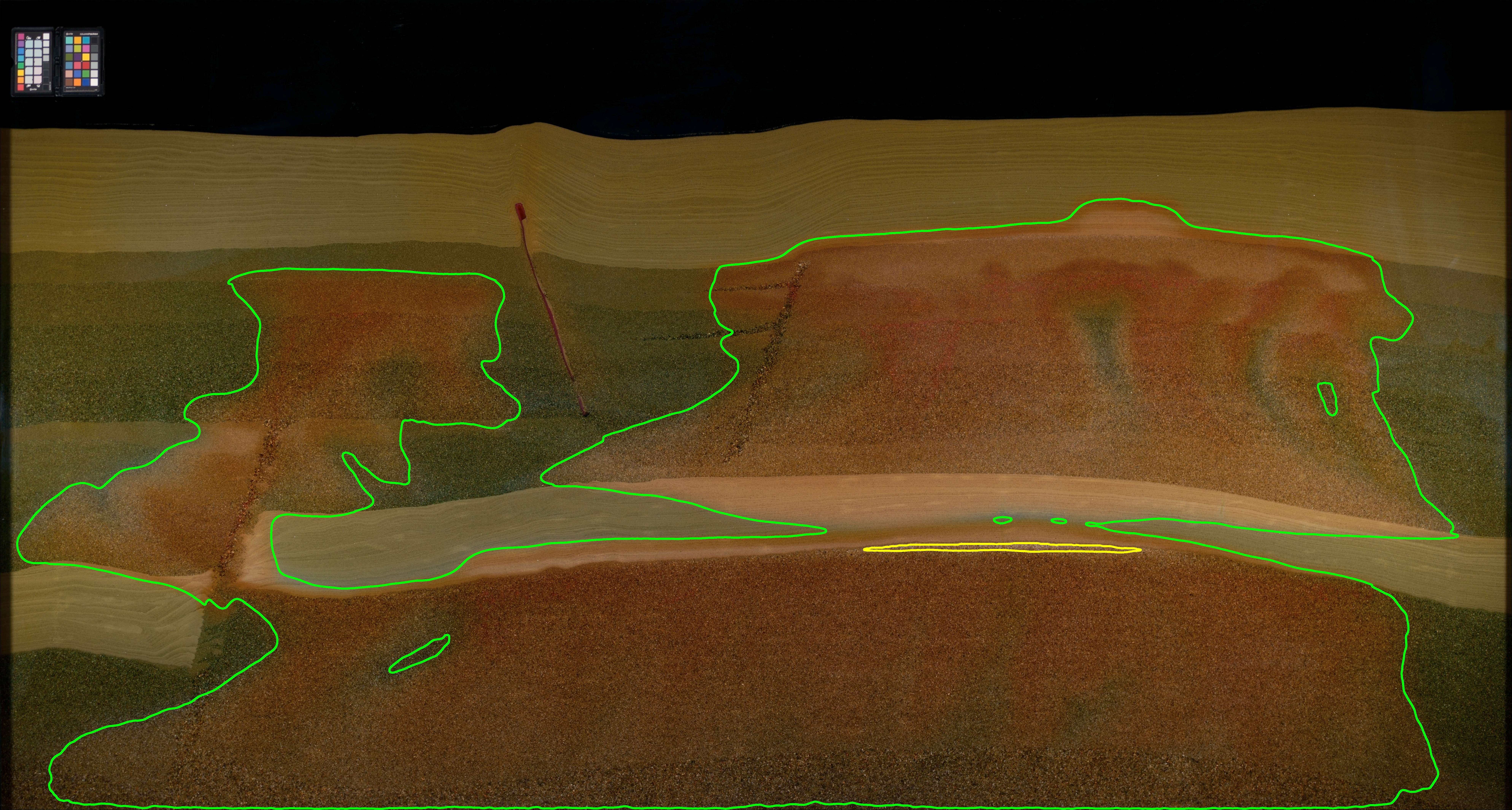} 
\caption{5 days}
\label{co2experimentc}
\end{subfigure}
\caption{Photographs from the CO$_2$ injection experiment (C1) showing contours of dissolved CO$_2$ (green) and gaseous CO$_2$ (yellow). a) Injection into the lower well only. b) Injection into both the lower and upper wells. c) Post-injection phase after both wells were shut down}
\label{co2experiment} 
\end{figure}

Figure~\ref{co2experiment} displays a subset of the experimental data from the CO$_2$ injection study, which is described in full detail by \cite{ferno2024}. Among the five experimental runs (C1-C5) conducted under nearly identical operational conditions, the results presented here correspond to run C2 (\cite{ferno2024}). In this run, CO$_2$ was injected at a rate of 10 ml/min under standard conditions (equivalent to 1.67$\times 10^{-7}$ m$^3$/s). Injection into the lower well (see Figure~\ref{ff_rig} for location) lasted for 5 hours and 5 minutes. Injection into the upper well began 2 hours and 15 minutes after the lower well injection started and continued for 2 hours and 50 minutes. Both wells were shut down simultaneously after 5 hours and 5 minutes of total injection time. The final image in the dataset was captured five days after the start of the experiment, marking the end of the experiment.

High-resolution photographs of the CO$_2$ migration define the data acquisition. The optical images have been processed using DarSIA~(\cite{DarSIA}), an open-source Python-based image analysis toolbox for Darcy-scale images. This tool enables the generation of segmented images, which are used to compare the simulation results submitted by various research groups to the five experimental runs as discussed in \cite{Flemisch2024}. The segmentation identifies the spatial distribution of the different fluid phases: gaseous CO$_2$, dissolved CO$_2$, and pure water. This segmented data also build the foundation for the history matching presented herein.

\subsection{Reservoir simulator}
The numerical simulations corresponding to the CO$_2$ injection experiments are performed using the open-source simulator OPM Flow (\cite{RASMUSSEN2021159}). OPM Flow is a reservoir simulator that supports industry-standard input and output formats, making it suitable for both research and practical applications. In this study, we briefly describe the mathematical model for GCS as implemented in OPM Flow, and provide references for readers seeking more detailed information. It is worth noting that this implementation has demonstrated strong performance in recent benchmark studies, including the 11th Society of Petroleum Engineers Comparative Solution Project (\cite{nordbotten2025}), which was inspired by the FluidFlower experimental setup.

\subsubsection{Mathematical model}\label{subsecmodel}
The OPM Flow simulator includes a dedicated module for CO$_2$ storage applications. When this option is enabled, the simulator internally computes fluid properties such as density, viscosity, and the solubility limit of CO$_2$ in brine. These properties are calculated as functions of pressure, temperature, and composition using analytical correlations and models from the literature, rather than relying on interpolation from tabulated data. Internally, these compositional properties are converted to their black-oil equivalents, allowing the simulator to retain the computational efficiency and robustness of a black-oil formulation while achieving the accuracy typically associated with compositional models. A detailed description of the CO$_2$ storage module in OPM Flow is provided in \cite{Sandve2021}. In this study, we adopt a simplified version of the CO$_2$ storage model tailored to the FluidFlower experiments. The formulation and notation used follow the conventions presented in \cite{10.2118/218015-PA}.

The pore space in the reservoir is occupied by a two-component, two-phase fluid system, consisting of water and CO$_2$ as the components ($i\in\left\{\text{H}_2\text{O}, \text{CO}_2 \right\}$), and liquid and gas as the phases ($\alpha\in\left\{l, g \right\}$). The two-phase extended Darcy's law for phase $\alpha$ is written as:
\begin{equation}
\mathbf{u}_\alpha=-\frac{k_{r,\alpha}\mathbf{k}}{\mu_\alpha}\left( \nabla p_\alpha - \rho_\alpha\mathbf{g} \right),
\end{equation}
where $\mathbf{u}_\alpha$ is the flux discharge per unit area [m/s], $k_{r,\alpha}$ the relative permeability [-], $\mu_\alpha$ the dynamic viscosity [Pa$\cdot$s], $p_\alpha$ the pressure [Pa], $\rho_\alpha$ the density [kg/m$^3$], $\mathbf{g}$ the gravity (9.81 [m/s$^2$]), and $\mathbf{k}$ a symmetric tensor of rank 2 for the rock permeability [m$^2$]. The component mass conservation is written as:
\begin{equation}
\sum_{\alpha=l,g}{\left\{\frac{\partial}{\partial t}\left[\rho_\alpha\phi s_\alpha\chi^i_\alpha \right]+\nabla\cdot\rho_\alpha\left[ \mathbf{u}_\alpha\chi^i_\alpha-\left(s_\alpha\phi D_\alpha+E\left\Vert\mathbf{u}_\alpha\right\Vert\right)\nabla\chi^i_\alpha \right] \right\}=0},
\end{equation}
where $\phi$ is the porosity [-], $s_\alpha$ the saturation [-], $\chi^i_\alpha$ the component mass fraction in phase $\alpha$ [-], $D_\alpha$ the molecular diffusion [m$^2$/s], $E$ the isotropic dispersion coefficient [m], and $\|\cdot\|$ denotes the Euclidean norm. The phase partitioning of each component is defined according to \cite{SPYCHER20033015}. Under room temperature and pressure conditions, water vaporization is minimal. Therefore, in the simulations presented in this study, vaporization of water into the gas phase ($\chi^{\text{H}_2\text{O}}_g$) is neglected. The phase saturations, pressures, and components fulfill the following conditions:
\begin{equation}
p_g-p_l=p_c, \quad s_l+s_g=1,\quad \text{and} \quad \sum_{i=\text{H}_2\text{O}, \text{CO}_2}\chi^i_\alpha=1\quad \alpha\in\left\{l, g \right\},
\end{equation}
where $p_c(s_l)$ is the capillary pressure [Pa]. In this study, we use the Brooks-Corey functions, where these relationships are given as a function of the effective saturations $s_\alpha^*$:
\begin{equation}\label{satfuncs}
s_\alpha^*=\max\left(\frac{s_\alpha-s_{\alpha,i}}{1-s_{\alpha,i}},0\right),\quad k_{r,\alpha}=(s_\alpha^*)^{n_\alpha},\quad \text{and}\quad p_c=p_e(s_l^*)^{-\tfrac{1}{n}},
\end{equation}
where $s_{\alpha,i}$ is the saturation below which the phase is immobile [-], $p_e$ the entry pressure [Pa], and $n_\alpha$ and $n$ fitting coefficients [-]. Although OPM Flow supports hysteresis modeling, it is intentionally omitted in the simulations presented in this study as a simplification choice to reduce model complexity.

The PVT (pressure, volume, temperature) properties such as densities and viscosities as a function of pressure, temperature, and composition are computed internally by using analytical correlations and models from the literature (\cite{coolprops}). We refer to the OPM Flow manual for details on these models (\cite{flowmanual}).

\subsubsection{Spatial characterization}\label{sec:grid}
\begin{figure}[h!]
\centering
\includegraphics[width=\textwidth]{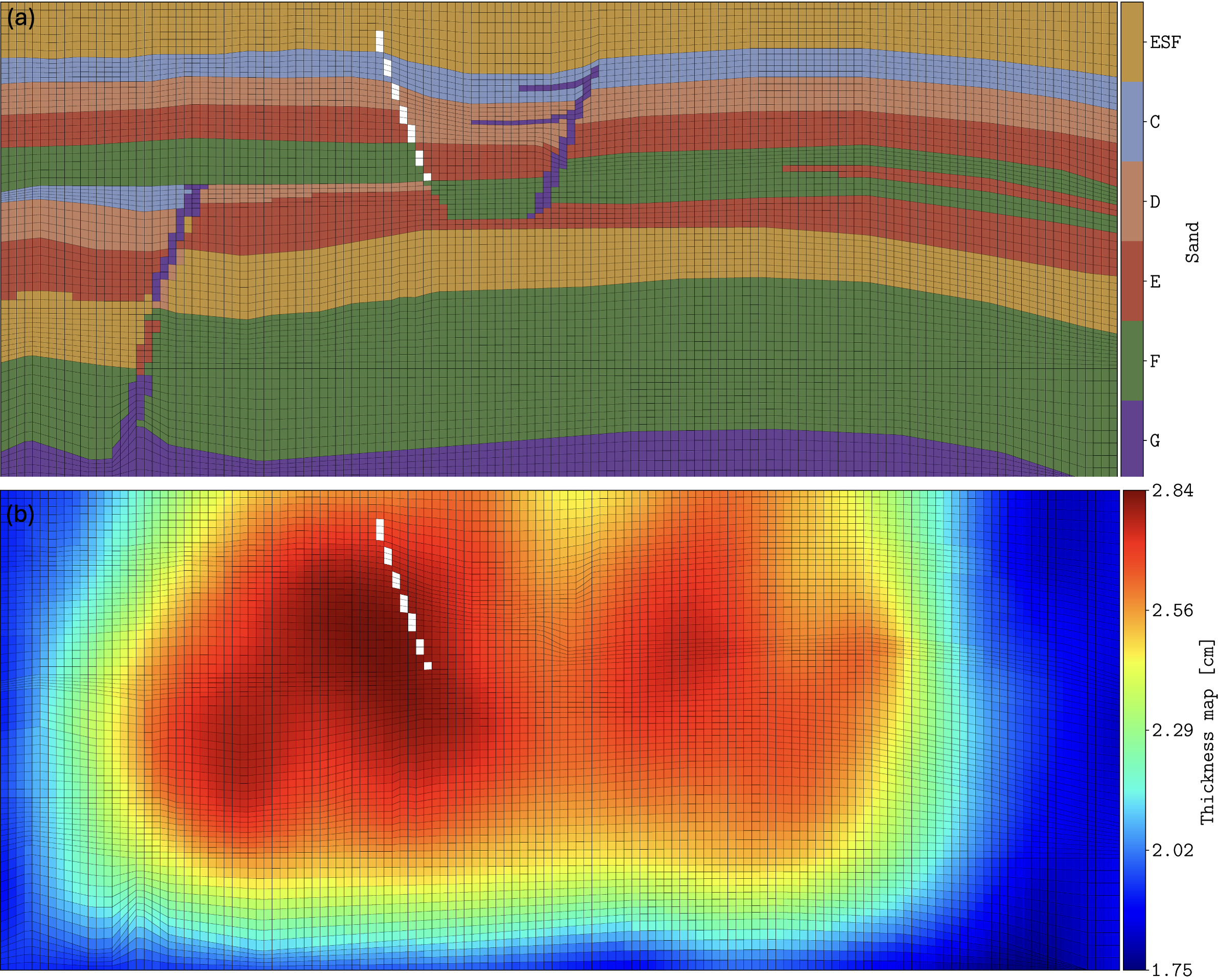}
\caption{(a) Corner-point grid representation of the FluidFlower geological model. (b) Measured thickness map of the FluidFlower}
\label{ff_grid}
\end{figure}

A grid representing the different facies in the FluidFlower rig is required as input to the simulator. Cartesian grids offer computational efficiency due to the use of the two-point flux approximation in the discretization scheme (\cite{mrst}). However, accurately representing the interfaces between sand layers with Cartesian grids requires a very fine spatial resolution, which can increase computational cost. Unstructured grids provide greater flexibility by allowing the grid to conform to complex geological features such as sand layers and faults. This adaptability comes with increased complexity in the discretization matrix and may introduce grid-orientation effects when using two-point flux approximation (\cite{Klemetsdal2017}). Corner-point grids are widely used in subsurface simulations because they balance geometric flexibility with computational efficiency (\cite{ponting1989}). These grids are defined by vertical pillars and horizontal lines connecting the pillars. For the simulations presented in this study, we use the corner-point grid shown in Figure \ref{ff_grid}, which allows for accurate representation of geological structures while maintaining compatibility with the numerical methods employed. We refer to \cite{gridorien2025} for an extensive study on grid-orientation effects and discretization methods on the FluidFlower geometry.

The computational grid in Figure~\ref{ff_grid} comprises 140 elements in the horizontal dimension, each with a uniform size of 2 cm, and 69 elements in the vertical dimension, where cell height varies from a minimum of 4.7$\times 10^{-3}$ cm to a maximum of 3.11 cm, with an average value of 1.74 cm. This configuration results in a total of 9,960 grid cells, of which 9,627 are active. The remaining 333 cells are inactive: 18 correspond to the sealed fault (see Figure~\ref{ff_rig}), while 315 are located in the lower-right corner of the grid, where the horizon of facies G coincides with facies F. The grid itself is defined with a constant thickness equal to the minimum value from the measured thickness map. Variations are incorporated indirectly by adjusting the pore volumes and transmissibilities of the cells.

\subsection{Modeling decisions and comparison metrics}
This study is closely aligned with the special issue on the ``FluidFlower validation benchmark study"~(\cite{nordbotten2024fluidflower}), which includes 14 research papers that contribute valuable knowledge relevant to the current work. This work in particular complements and reinforces the collective understanding of the benchmark, offering additional validation and perspective. Additionally, results from the 11th SPE Comparative Solution Project (\cite{nordbotten2025}), a benchmark study inspired by FluidFlower, are also pertinent to this research. This subsection highlights key findings from both contributions that informed our modeling decisions, particularly regarding grid type and resolution.

As part of the later discussion, our history matched results will be put into direct comparison with not only the experimental data itself, but also the other modeling contributions to the model validation benchmark~(\cite{Flemisch2024}). One of the metrics employed in this comparison is the Wasserstein distance, also referred to as the Earth mover’s distance or optimal transport distance (\cite{villani2008optimal}). For two distributions $\rho_\mathrm{A}:\Omega \rightarrow \mathbb{R}_+$ and $\rho_\mathrm{B}:\Omega \rightarrow \mathbb{R}_+$ defined over the space $\Omega$ and of same measure, i.e., $\int_\Omega \rho_\mathrm{A} \,dV = \int_\Omega \rho_\mathrm{B} \, dV$, their Wasserstein distance is defined through a nonlinear minimization problem originating from optimal transport, also called Beckmann problem (\cite{santambrogio2015optimal}):
\begin{align}
    W^1(\rho_\mathrm{A}, \rho_\mathrm{B}):= \mathrm{min} \,\left\{\left.\int_\Omega |\bm{q}|\, dV \ \right|\ \nabla \cdot \bm{q} = \rho_\mathrm{A} - \rho_\mathrm{B},\ \left.\bm{q}\cdot \bm{n}\right|_{\partial\Omega} = 0 \right\}.
\end{align}
In words, this metric determines the weakest flux field that transports the one distribution into another. The use of the $L^1$ norm localizes the transport, resulting in a shortest path. With this, it provides an objective measure for data with inherent transportation character and is convenient for model-experiment comparisons in flow in porous media (\cite{Both2024}). Among all nine submitted simulation results, the results from CSIRO show the closest agreement with the experimental data in terms of the Wasserstein distance (\cite{Green2024}); their results will be later explicitly highlighted in the discussion and plots of the HM results. 

In an associated study, \cite{Salgado2023} manually calibrated model parameters using experimental data from a smaller-scale setup and used these parameters to simulate the FluidFlower system. One key finding from this study is the high sensitivity of the system to variations in permeability and capillary pressure. To evaluate model performance, the Wasserstein distance was computed relative to the experimental results. Among the tested configurations, Model 1 showed the closest agreement with the experimental data; again, the corresponding results will be explicitly highlighted in the discussion and plots of the HM results. In our paper, Model 1 is referred to as MIT\_M1.

Several studies of the FluidFlower special issue discuss the sensitivity of simulation results to model choices and parameter variations. One study investigates the roles of hysteresis and molecular diffusion, finding that hysteresis contributes minimally, whereas enhanced diffusion significantly improves dissolution trapping (\cite{Wang2024}). Another contribution compares discretization techniques, evaluating Cartesian versus unstructured grids and contrasting two-point with multi-point flux approximations (\cite{Wapperom2024}). A further contribution uses ensemble simulations to assess uncertainties in rock and gas relative permeability, revealing a notable sensitivity to the image segmentation threshold, particularly when comparing the assigned value in the FluidFlower benchmark study of 0.1 kg/m$^3$ (CO$_2$ concentration in the liquid phase) with an alternative of 0.05 kg/m$^3$ (\cite{Jammoul2024}). Additionally, permeability field heterogeneity is analyzed through HM at each grid cell, incorporating both tracer and CO$_2$ concentration data (\cite{Tian2024}).

\begin{figure}[h!]
\centering
\includegraphics[width=\textwidth]{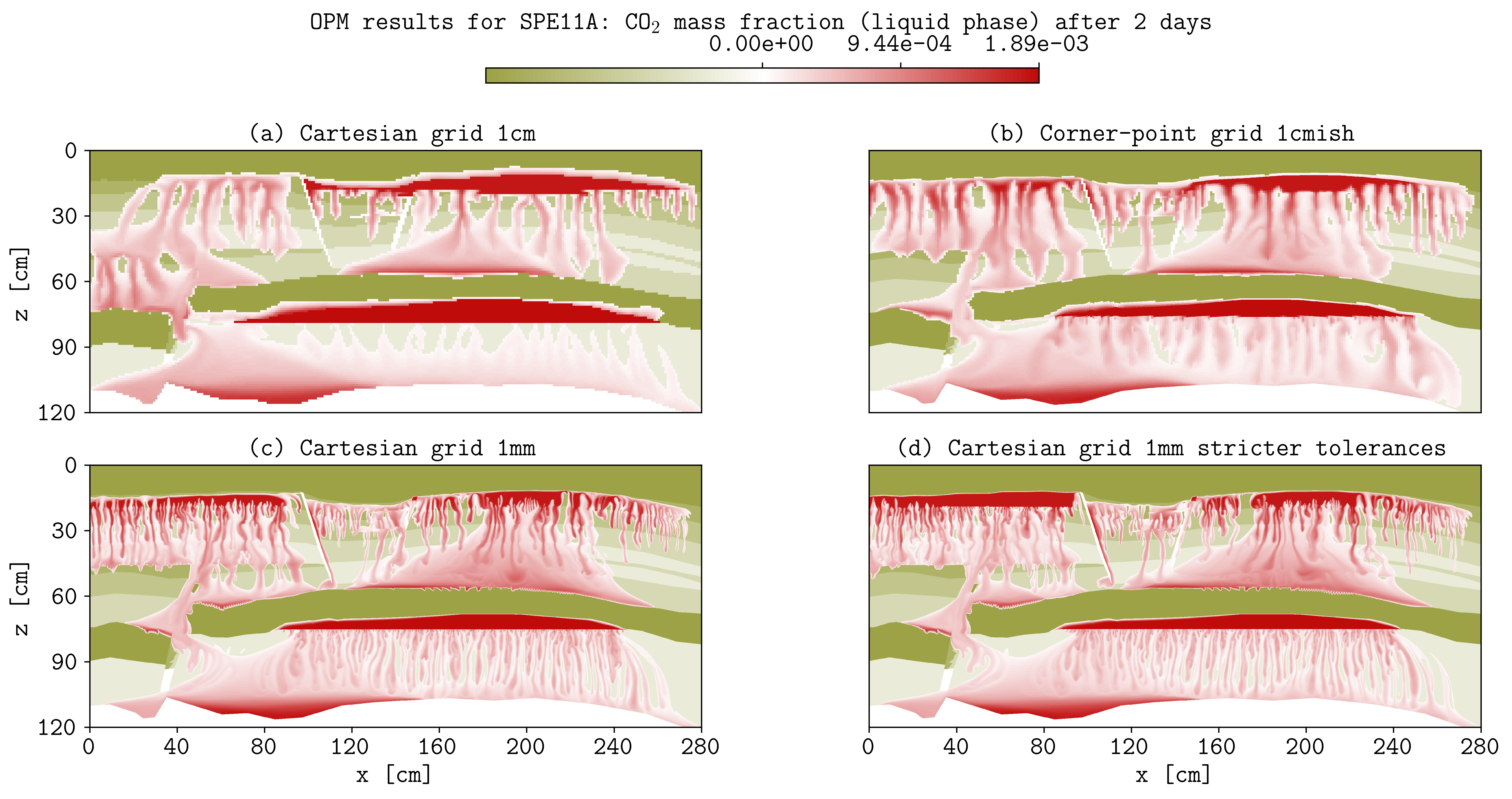}
\caption{Spatial distribution of dissolved CO$_2$ from the OPM simulation results in case SPE11a, which closely resembles the FluidFlower experimental system}
\label{spe11a_maps}
\end{figure}

One of the key findings from the SPE11 benchmark study is that human errors appear to be the primary source of variability in the simulation results (\cite{nordbotten2025}). This underscores the importance of open-source code, which enables independent validation and verification by the broader research community. Another relevant contribution to the present study is the comparison of grid structures and convergence criteria, provided by the OPM team. Figure \ref{spe11a_maps} presents spatial maps of dissolved CO$_2$ after two days of simulation. For details on the simulation setup, configuration files, and additional results, refer to the documentation of the \texttt{pyopmspe11} tool (\cite{Landa-Marbán2025}).

Figure \ref{spe11a_maps}a, corresponding to SPE11 simulation results close to the considered FluidFlower setup, illustrates that in simulations using 1 cm Cartesian grids, convective mixing ceases in the lower storage unit, an artifact attributed to this grid configuration (\cite{Flemisch2024}). Figure \ref{spe11a_maps}b presents results obtained with a corner-point grid, which show a better agreement than Figure \ref{spe11a_maps}a (Cartesian grid) with the fine-scale simulations depicted in Figures \ref{spe11a_maps}c and \ref{spe11a_maps}d. The difference between these two fine-scale cases lies in the solver tolerances, with \ref{spe11a_maps}d having stricter tolerances (refer to the online documentation of the \texttt{pyopmspe11} tool (\cite{Landa-Marbán2025}) for the values of these tolerances). Tightening the solver tolerances improves mass conservation, where in this system the theoretical injected mass is 4.59$\times 10^{-3}$ kg, while the injected mass in Figure \ref{spe11a_maps}c is 4.43$\times 10^{-3}$ kg (3.5\% error), and in Figure \ref{spe11a_maps}d is 4.58$\times 10^{-3}$ kg (0.2\% error). However, this improvement comes at the cost of increased simulation time, rising from approximately 17 days in Figure \ref{spe11a_maps}c to around 55 days in Figure \ref{spe11a_maps}d.

Overall, the results presented in Figure \ref{spe11a_maps} demonstrate that simulation outcomes are highly sensitive to grid type and resolution, as well as to solver tolerances. While high-resolution grids combined with strict solver settings are desirable for accuracy, they are impractical for HM studies, which typically require numerous simulation runs. Therefore, for our HM analysis, we adopt the coarser corner-point grid shown in Figure \ref{ff_grid}, with a resolution of approximately 2 cm (2 cm in the x-direction and around 2 cm in the z direction). This grid is sufficiently coarse to avoid mass conservation issues when using default solver tolerances. The choice of this grid is motivated by three factors: it adequately captures the heterogeneity of the fault zone, it offers acceptable simulation times (approximately 8 minutes per run in serial, 2 minutes in parallel using eight cores), and it aligns with the resolution used in the FluidFlower benchmark study for computing the Wasserstein distance, which was based on a 2 cm Cartesian grid. Note that since the simulation results are sensitive to grid size, the history-matched parameters will be grid-dependent and cannot be directly applied to other grids and configurations.

\subsection{History matching}
Data assimilation refers to the process of integrating measured data into a model to enhance its predictive accuracy. HM is a specific form of data assimilation in which reservoir parameters are adjusted to align simulation outputs with observed data. This approach typically involves generating an ensemble of reservoir models by sampling from distributions of uncertain parameters. The ensemble framework helps quantify and reduce uncertainty in both model parameters and measured data. Widely used algorithms for HM include the Ensemble Kalman Filter (\cite{evensen2003}) and the Ensemble Smoother (\cite{emerick2013ensemble}), which update model states and parameters based on observed measurements while preserving computational efficiency. Refer to \cite{Tian2024} for an ensemble-base HM for the FluidFlower system.

An alternative approach to Ensemble Kalman Filter and Ensemble Smoother that has gained increasing attention is the differential evolution method, a population-based optimization algorithm (\cite{Storn1997}). The differential evolution method is designed to iteratively improve a population of candidate solutions and is particularly well suited for complex optimization problems. It performs effectively in scenarios involving non-differentiable, non-linear, and multimodal objective functions, making it a robust choice for reservoir HM where the solution space is often irregular and high-dimensional. For our HM study, we use the differential evolution algorithm available via the \texttt{SciPy} library (\cite{Virtanen}) and is integrated using the Everest decision-making tool.

The HM algorithm requires a metric to quantify the mismatch between observed data and simulation results and guide the iterative improvement of the tuning parameters. For this purpose, we utilize the Wasserstein distance, which also built the foundation for the dense data comparison of the benchmark~(\cite{Flemisch2024}); we leverage the same Python implementation of its computation provided in the associated data repositories, also used in~\cite{Salgado2023}. The comparison is performed on spatial maps at 24, 48, 72, 96, and 120 hours, matching the time points reported in \cite{Flemisch2024}. The total cost, provided to the HM algorithm consists of the sum of the five distinct Wasserstein distance values. 

\subsection{Integration}
In this subsection, we describe the followed steps to conduct the HM. In order to assess the efficacy, we evaluate the distance to the experimental benchmark data and compare it against the other submissions to the model validation benchmark~(\cite{Flemisch2024}).

As previously explained, the FluidFlower system presents significant modeling challenges due to its high sensitivity to input parameters, grid configuration, and simulation settings. In this study, the parameters considered for HM include permeability, entry pressure, and residual saturations for each of the six sand layers, resulting in a total of 24 parameters. The temperature is set to 20$^{\circ}$C, and the pressure on the top boundary (at 1.2 m) is set to 104,900 Pa to match the pressure in the sensors. Mechanical dispersion is neglected to reduce the parameter space, which is a justified model simplification due to the coarse grid resolution. In addition, porosity values are not included in the HM process; instead, they are set equal to those used in Model MIT\_M1 from \cite{Salgado2023}, which are selected from publish data in similar silica sands. This decision is based on the observation that simulation results are extremely sensitive to porosity variations, which posed difficulties for the considered HM algorithm during preliminary testing.


To illustrate the scale of the HM problem, consider a simplified scenario where each of the 24 parameters (permeability, entry pressure, and residual saturations in the six sands) is sampled at two values (minimum and maximum), resulting in $2^{24}=16,777,216$ possible combinations. Expanding this to three values per parameter (e.g., minimum, middle, and maximum) increases the number of combinations to over $10^{11}$. In practice, the parameter space is even larger, e.g., this study considers 50 possible values per permeability interval, 20 for entry pressures, and between 15 and 4 for residual saturations (see Table \ref{iniparameters}), resulting in $6\times10^{28}$ possible combinations. Given this vast search space, the success of HM depends not only on the choice of algorithm but also on additional factors such as the initial parameter values and the selected random seed, which introduces an element of stochasticity and, to some extent, luck (\cite{rescher2021luck}).

Table \ref{iniparameters} summarizes the initial parameter range and number of samples. These parameter ranges and number of samples are set based on our preliminary testing. To conduct the HM, we adopt a sequential approach commonly used in the oil and gas industry, where different groups of parameters are adjusted in successive steps (\cite{YIN2011116}). In the first iteration, the parameters subject to HM are the permeabilities, residual saturations, and entry pressures for sands two through five. This choice is motivated by the fact that these formations constitute the primary storage sands, and simulation outcomes are highly sensitive to variations in these parameters. In the second iteration, the same set of parameters is adjusted, but only for sands one and six, which were not included in the first iteration. Since $s_{r,g}$ for sand five was 0.05 after the first iteration, then the value for $s_{r,g}$ for sand six was set to 0 and not HM. In the final iteration, a second adjustment of permeability, residual saturations, and entry pressure for sand five is performed. The rationale for revisiting sand five in the last step is that since CO$_2$ is injected into the lower unit, which consists primarily of sand five, its properties play a critical role in controlling upward leakage and density driven fingers. In addition, a constraint is imposed throughout the HM workflow to ensure monotonicity in the parameter values, e.g., permeability must increase with increasing grain size.
\begin{table}[h]
\centering
\caption{Initial model parameters [min, max, \# equidistance samples]}\label{iniparameters}
\begin{tabular}{@{}lrrrrr@{}}		
\toprule
Id & Sand&$k$ [D]&$s_{r,w}$ [-]&  $s_{r,g}$ [-] & $p_e$ [Pa]\\
\midrule
1 & ESF	&[50, 150, 50] &[0.32,  0.35, 4]& [0.20, 0.35, 15]& [1000, 2000, 20]\\
2 & C	&[50,  200, 50]  &[0.05,  0.35, 10]& [0.05,  0.35, 10]& [500, 1000, 20]\\
3 & D	&[100, 500, 50]  &[0.05, 0.32, 10]& [0.05, 0.32, 10]& [150, 600, 20]\\
4 & E	&[300, 1300, 50] &[0.05, 0.3, 10]& [0.05, 0.30, 10]& [100, 400, 20]\\
5 & F	&[1300, 3000, 50] &[0.05, 0.28, 10]& [0.05, 0.28, 10]& [50, 200, 20]\\
6 & G	&[2500, 5000, 50] &[0.05, 0.26, 10]& -\footnotemark[1]& [0, 170, 20]\\
\bottomrule
\end{tabular}
\\$^1$ Set to 0.\quad\quad\quad\quad\quad\quad\quad\quad\quad\quad\quad\quad\quad\quad\quad\quad\quad\quad\quad\quad\quad\quad\quad\quad\quad\quad\quad\quad\quad\quad\quad\quad
\end{table}
The configuration files containing all necessary details to reproduce this study are available at \url{https://github.com/cssr-tools/pofff/paper}. The configuration file used for the sequential HM approach (first iteration) is provided in Appendix \ref{secA1}. Simulations were executed on a local server equipped with 144 CPUs, and approximately 600,000 runs were performed. 

\section{Results and Discussion}\label{sec3}
In this section, we present the results of the image-based HM framework, when applied to the FluidFlower benchmark data to tune various uncertain material parameters as elaborated in the previous section. Table \ref{hmparameters} summarizes the resulting values of the HM parameters. 

\begin{table}[h]
\centering
\caption{Final model parameters after the HM}\label{hmparameters}
\begin{tabular}{@{}lrrrrrr@{}}		
\toprule
Id & Sand&$k$ [D]&$\phi$  [-]\footnotemark[1] &$s_{r,w}$ [-]&  $s_{r,g}$ [-] & $p_e$ [Pa]\\
\midrule
1 & ESF	&62	 	&0.37&0.34& 0.25& 1900\\
2 & C	&152	&0.38&0.32& 0.20& 950\\
3 & D	&428	&0.40&0.29& 0.08& 185\\
4 & E	&1120	&0.39&0.27& 0.06& 175\\
5 & F	&2014 	&0.39&0.26& 0.01& 170\\
6 & G	&2500	&0.42&0.17& 0& 163\\
\bottomrule
\end{tabular}
\\$^1$ Porosity values from model M1 in \cite{Salgado2023}.
\end{table}
%
%
\noindent Although the HM values are generally of the same order of magnitude as the measured values in Table \ref{parameters} (used by CSIRO) and the ones for MIT\_M1 (see \cite{Salgado2023}), some differences are observed. Notably, the entry pressure values obtained from the HM are consistently higher. This is expected, as coarse grids in numerical models lead to overestimation of entry pressure, since simulations in coarse grids underestimate CO$_2$ dissolution rates (\cite{su142215049}), while higher entry pressures leads to higher CO$_2$ dissolution rates (\cite{Martinez2016}). In addition, the CSIRO simulations used a constant thickness of 25 mm and the MIT\_M1 simulations employed a different thickness map derived from initial experimental data (see Fig. 2c in \cite{Salgado2023}), whereas our simulations used the final measured thickness map (Fig. \ref{ff_grid}b). The relative permeability and capillary pressure models also differ between the CSIRO (see Eqs. 8-11 in \cite{Green2024}), MIT\_M1 (see Subsection 3.2.1 in \cite{Salgado2023}), and our simulations (Eq. \ref{satfuncs}). The differences between thickness maps and saturation function models, combined with the use of distinct grid types and resolutions across the groups, contribute to the variation in parameter values.

\begin{figure}[h!]
\centering
\includegraphics[width=\textwidth]{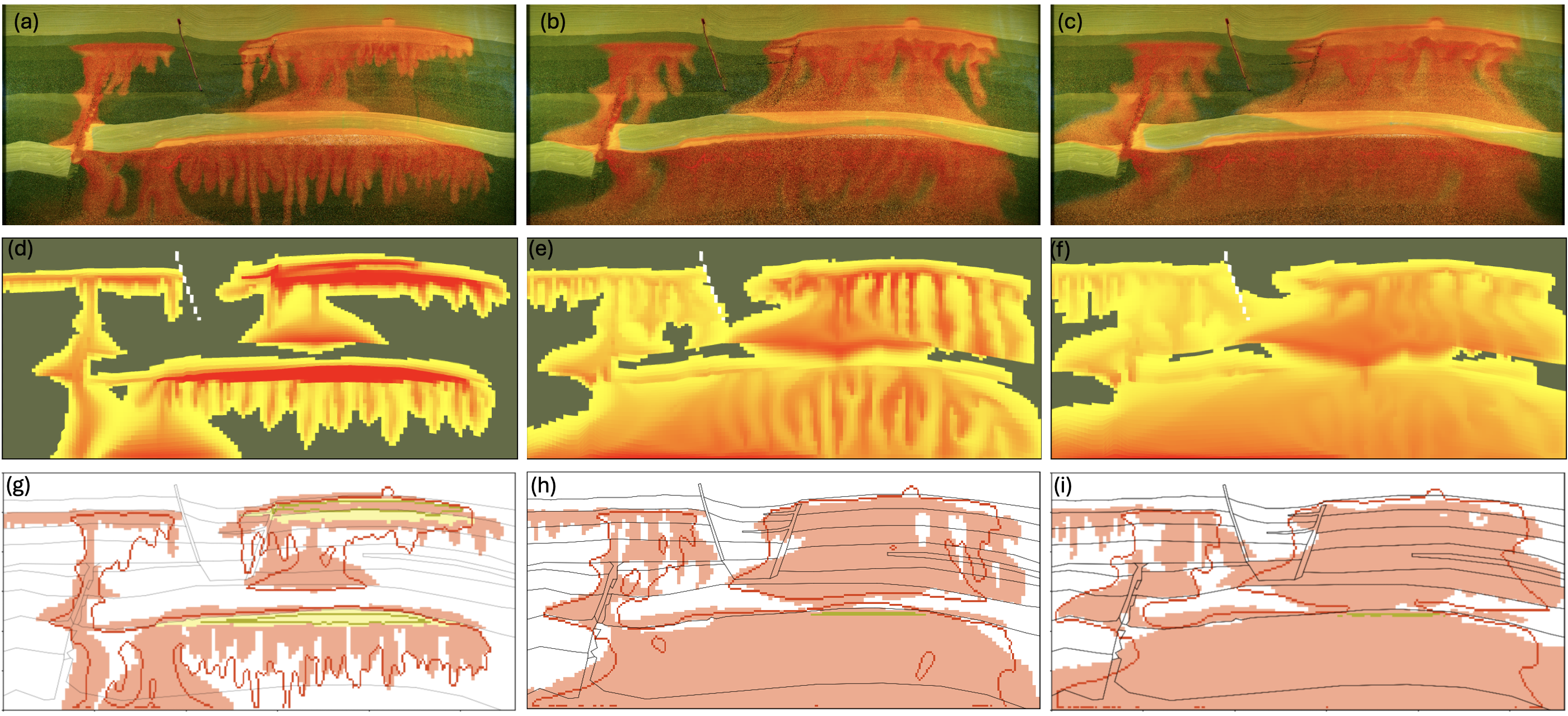}
\caption{(a-c) Photographs from the FluidFlower experiment C2 at one, three, and five days, respectively. (d-e) Spatial maps of CO$_2$ concentration from the corresponding simulation results. (f-g) Contour plots comparing experimental observations and segmented simulation outputs at the same time intervals}
\label{ff_maps}
\end{figure}

Figure \ref{ff_maps} presents spatial maps comparing the simulation results, using the history-matched parameter set, and the experimental data. Figures \ref{ff_maps}a-f show good agreement between the simulation results and the experimental data. However, the segmented simulation maps in Figures \ref{ff_maps}g-i exclude certain CO$_2$ regions, particularly in the upper-left area, resulting in the simulated CO$_2$ reaching the upper-left boundary. This discrepancy is related to the segmentation thresholds used in the benchmark study, where a gas saturation threshold of 0.01 and a CO$_2$ concentration threshold of 0.1 kg/m$^3$ were applied. These thresholds introduce additional sensitivity into the analysis, as also discussed by \cite{Jammoul2024}. In Appendix \ref{secA2}, simulation results using a lower threshold of 0.05 for the CO$_2$ concentration are shown. To mitigate the impact of threshold selection, ongoing research is focused on improving the image processing workflow utilizing regression instead of segmentation (\cite{folkvord2025laboratory}). The goal is to generate continuous maps of CO$_2$ mass directly from experimental images, enabling more robust comparisons with simulation results.

\noindent To further evaluate the quality of the HM simulation results and to place them in context with the FluidFlower benchmark figures~(\cite{Flemisch2024}), we compute the Wasserstein distance between all experimental datasets (experiments C1 through C5) and the corresponding simulation outputs. As the HM workflow explicitly aims at minimizing the Wasserstein distance, it is expected that the HM yields to the overall best agreement in this metric. In addition to dense data comparisons, the benchmark study also included  time-series and sparse data corresponding to effective quantities in pre-defined boxes~(\cite{Nordbotten:Manual:2022,Flemisch2024}). As these were not included in the HM workflow, the comparison of the time-series and sparse data is of high relevance. For comparison, we highlight the individual result from CSIRO, being the participant with the closest results to the experimental data, and include  the result MIT\_M1 from \cite{Salgado2023}, where model parameters were set from published data based on the measurements of the average grain sizes, and the capillary pressure curve for sand D was manually HM. Our HM result is labeled CSSR, representing the research center that supported this study (see Acknowledgments).

\begin{figure}[h!]
\centering
\includegraphics[width=\textwidth]{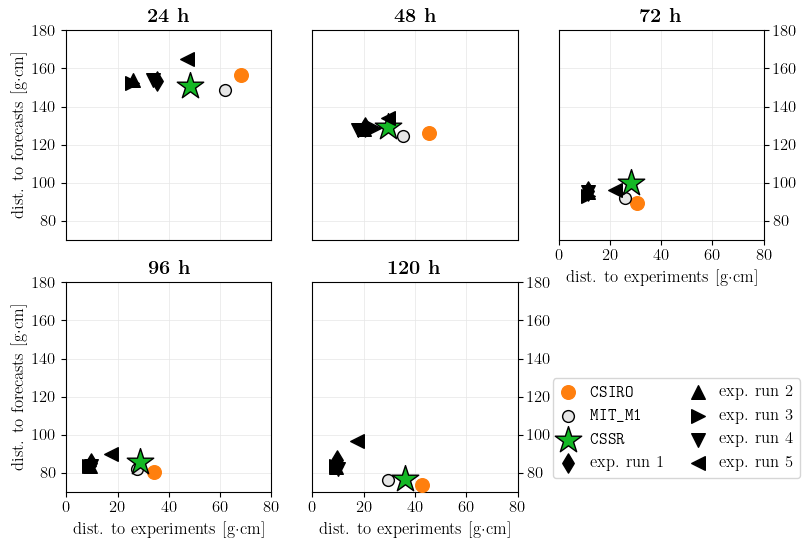}
\caption{Wasserstein distances computed between simulation results and experimental data (experiments C1-C5), as well as between simulation forecasts. Forecast comparisons include only results submitted by participants in the FluidFlower benchmark study}
\label{ff_emd}
\end{figure}

\begin{figure}[h!]
\centering
\includegraphics[width=\textwidth]{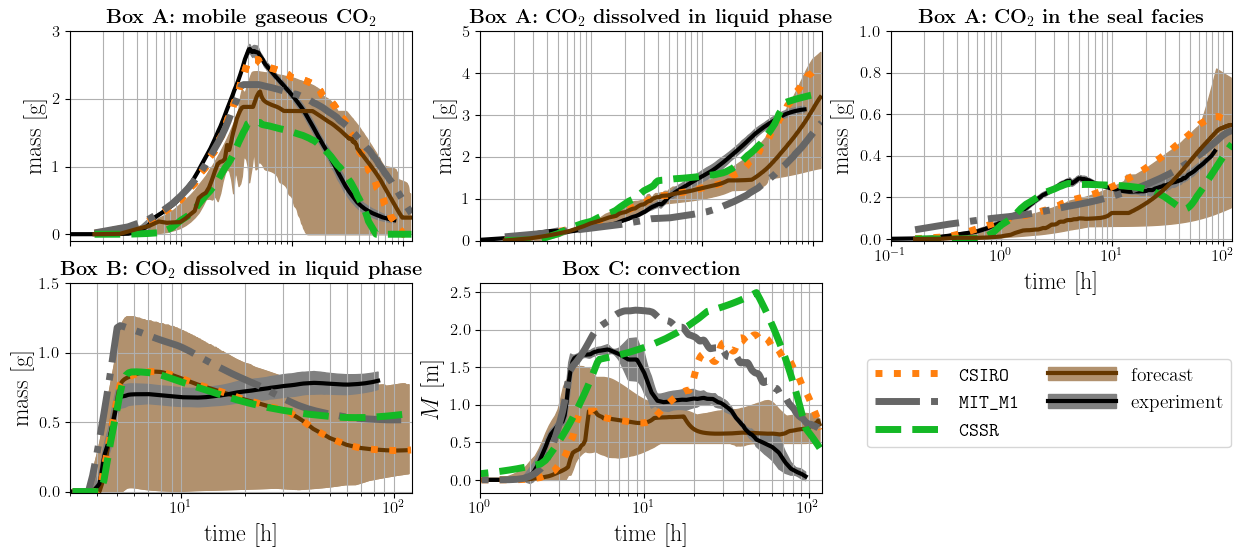}
\caption{Comparison between simulation forecasts and experimental observations for the temporal evolution of box quantities. The brown line represents the median of the simulation results submitted by benchmark participants, with the shaded pale brown region indicating the interquartile range (from the first to the third quartile). The black line shows the mean of the experimental measurements, and the surrounding gray region reflects the variability expressed as one standard deviation}
\label{ff_timeseries}
\end{figure}

\begin{figure}[h!]
\centering
\includegraphics[width=\textwidth]{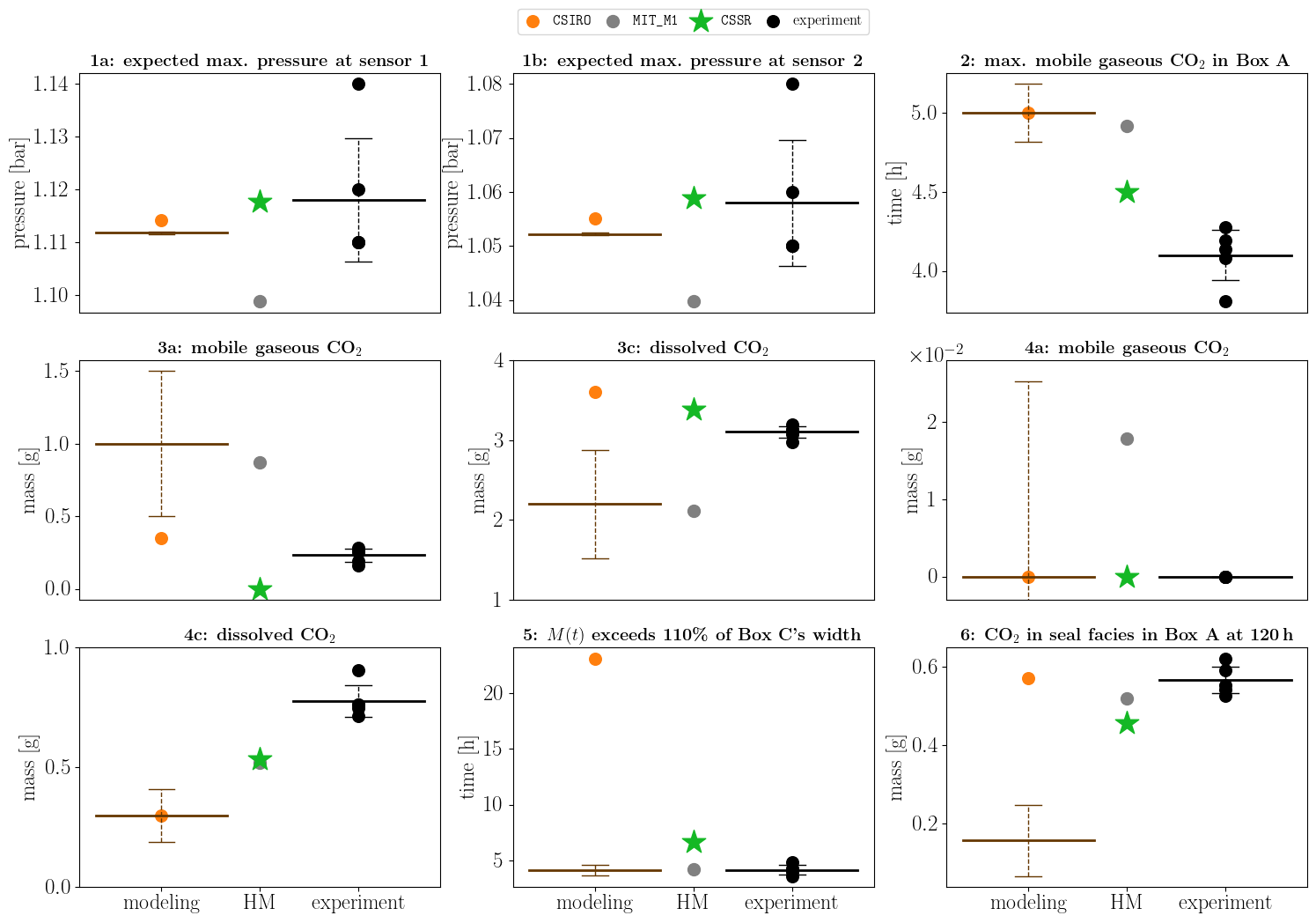}
\caption{Comparison of sparse data reported by benchmark participants with experimental observations. For the simulation results, the horizontal brown line represents the median, while the dashed vertical brown line indicates $\pm$ the median of all reported P50 values for the standard deviation. The middle x-axis (HM) displays the results from \cite{Salgado2023} (MIT\_M1) and from this study (CSSR). For the experimental data, black circles represent individual measurements from each run, the horizontal black line shows the mean, and the dashed vertical black line denotes $\pm$ one standard deviation}
\label{ff_sparse}
\end{figure}

Figure \ref{ff_emd} displays the spreading of the Wasserstein distance and shows that all experimental results yield Wasserstein distances below 50 g$\cdot$cm, with the closest distance being approximately 10 g$\cdot$cm. The results from CSIRO demonstrate excellent agreement with the experimental data under this metric, especially considering that benchmark participants did not have access to the CO$_2$ experimental data prior to submission. The calibrated model from \cite{Salgado2023} (MIT\_M1) also performs well, supporting the available published data for the different grain sizes and the effectiveness of manual parameter tuning. As expected, our history-matched result (CSSR) shows similarly close agreement with the experimental data, highlighting the quality of the HM approach used in this study. 

The time-series data is displayed in Figure~\ref{ff_timeseries}. For the temporal evolution of these box quantities, the results from CSIRO, MIT\_M1, and CSSR follow similar trends across all metrics, with the exception of box c, which captures convective behavior, defined as the integral of the magnitude of the gradient in relative concentration of dissolved CO$_2$. This quantity is particularly challenging to capture, as previously discussed in \cite{Flemisch2024}. 

Finally, Figure \ref{ff_sparse} displays the comparison of the sparse data. It shows that the closest match to the experimental mean varies depending on the specific sparse data quantity. For example, CSSR performs best for quantity 2, CSIRO for quantity 3a, and MIT\_M1 for quantity 5. To quantify these comparisons, we compute the error with respect to the experimental data for each sparse quantity across all datasets as follows:
\begin{equation}\label{errorg}
\text{error}_g=\frac{100}{6}\sum_{i\in\left\{ 2,3a,3c,3d,4c,6\right\} }\left|\frac{g_i-\text{exp}_i}{\text{exp}_i}\right|,\; g\in \left\{ \text{CSIRO, MIT\_M1, CSSR} \right\}
\end{equation}
where the quantity 2 is the time of max CO$_2$ mobile free phase in box a [s], 3a the CO$_2$ mobile free phase in box a at 72h [g], 3c the CO$_2$ dissolved in water in box a at 72h [g], 3d the CO$_2$ in the seal in box a at 72h [g], 4c the CO$_2$ dissolved in water in box b at 72h [g], and 6 the total mass of CO$_2$ in the top seal facies [g]. The results are scaled by a factor of 100/6 to bring the values into the same order of magnitude as the Wasserstein distance, see Table~\ref{errors}. Note that pressure values (quantities 1a and 1b) are excluded from this analysis, as our simulations impose a top boundary condition that matches the experimental mean. Additionally, box C (quantity 5) is omitted due to the modeling challenges previously discussed in \cite{Flemisch2024}.

\begin{table}[h]
\centering
\caption{Sparse data and Wasserstein distance (WD) comparison between experimental and simulation results}\label{errors}
\begin{tabular}{@{}lrrrrrrrrr@{}}		
\toprule
Data source & \# cells & 2 (s) & 3a (g) & 3c (g) & 3d (g) & 4c (g) & 6 (g) & error$_\text{g}$\footnotemark[1] & WD (g$\cdot$cm)\\
\midrule
Experiment & - & 14800 & 0.23 & 3.10 & 0.38 & 0.78 & 0.57 & - & 18.32 \\
MIT\_M1 & 151,402 & 17700 & 0.87 & 2.11 & 0.43 & 0.52 & 0.52 & 63.60 & 35.98 \\
CSIRO & 44,284 & 18000 & 0.35 & 3.60 & 0.57 & 0.30 & 0.57 & 33.38 & 44.21 \\
CSSR & 9,627 & 16200 & 0 & 3.38 & 0.28 & 0.53 & 0.46 & 32.85 & 34.06 \\
\bottomrule
\end{tabular}
\\$^1$ See Eq. \ref{errorg}.\quad\quad\quad\quad\quad\quad\quad\quad\quad\quad\quad\quad\quad\quad\quad\quad\quad\quad\quad\quad\quad\quad\quad\quad\quad\quad\quad\quad\quad\quad\quad\quad\quad\quad\quad\quad\quad\quad
\end{table}

\noindent From Table~\ref{errors}, we observe that the results obtained using the HM presented in this paper, CSSR, yield the lowest errors for both the sparse data and the Wasserstein distance metrics. It is noteworthy that the HM resulted in a 0 value for 3a (CO$_2$ mobile free phase in box a at 72h [g]), compared to the experimental value of 0.23. While this still deviates from the observed data, it represents a closer match than the MIT\_M1 prediction of 0.87. Table~\ref{errors} also shows the number of cells for each of the three simulation studies, with our grid having the less number of cells. The number of cells (grid size) for the MIT\_M1 was selected after doing a grid refinement study, where the grid size of 5 mm was needed to get accurate results (\cite{Salgado2023}). Additionally, the simulation time (in serial) for a single run using our coarse grid is approximately 8 minutes, which can be even less than two minutes running the simulator in parallel (the simulation times for CSIRO and MIT\_M1 are not available). This observation brings us back to the broader discussion on the trade-off between accuracy and computational efficiency, a balance that remains difficult to quantify, particularly when considering how to weight the errors in Table~\ref{errors} against the corresponding run times/number of cells.

\section{Conclusion and Outlook}\label{sec4}
In this work, we presented an image-based history matching workflow applied to the International FluidFlower benchmark data. In summary, the exercise enabled additional validation of the modeling of CO$_2$ at laboratory conditions.

HM relies on the ability to perform hundreds of simulations efficiently and reliably. Simulating the CO$_2$ injection experiment under room conditions using Darcy-scale permeability (as opposed to the more common milliDarcy range in reservoir applications) presents challenges for traditional reservoir simulators. Achieving sufficiently accurate solutions requires small time steps, which significantly increases simulation time when using fine grids on the order of millimeters. This, in turn, limits both the total number of simulations and the number of iterations feasible within the HM workflow.
To address this, we designed and tuned a coarse corner-point grid with a cell size of approximately 2 cm. This adjustment reduced the simulation time to around 8 minutes per run (2 minutes in parallel using eight cores), enabling the execution of hundreds of thousands of simulations on local servers within a few weeks. For this study, we utilized a local server equipped with 144 CPUs. After the HM process, the final simulation results show good agreement with the experimental CO$_2$ injection data. It would be desirable to further explore the predictive capability of the updated effective model against forecast data. Unfortunately, this validation is not possible given the lack of additional experimental scenarios in the original benchmark study. Nonetheless, the successful results of the present study reinforces the findings presented in the evaluation of the FluidFlower benchmark~(\cite{Flemisch2024}), which suggest that the system can be effectively described using standard flow equations, conventional saturation functions, and typical PVT properties for CO$_2$-brine mixtures.

The developed Python package for OPM Flow and FluidFlower, \texttt{pofff}, enables full reproducibility of the results presented in this study. Moreover, the input configuration file provides a flexible framework for exploring alternative approaches to HM using FluidFlower data. This includes testing different parameter distributions, applying various strategies for parameter splitting across HM steps, adjusting simulation settings to reduce computational time, and evaluating alternative models for saturation functions. It is important to note that simulation results are highly sensitive to changes in grid resolution, model formulations (such as the exponents in the relative permeability and capillary pressure relationships), and threshold values, including tolerances used for simulation segmentation. The history-matched parameters are thus not directly applicable to other setups. Given the vast space of possible model configurations (e.g., grid type and size, saturation models, and parameter values), \texttt{pofff} includes functionality for qualitative comparison of simulation outputs against both experimental data and the final results presented in this study (CSSR, see Table~\ref{errors}). Users are encouraged to contribute their configuration files to a dedicated folder within \texttt{pofff}, facilitating collaborative efforts toward improved matches and model refinement.

\paragraph{Author Contributions} All authors contributed to the study conception and design. Code implementation, numerical simulations, and analysis were performed by David Landa-Marb\'an, Tor H. Sandve, and Jakub W. Both. The first draft of the manuscript was written by David Landa-Marb\'an. All authors read and approved the final manuscript.

\paragraph{Data Availability} The simulator OPM Flow can be obtained at \url{https://opm-project.org}. Link to complete codes for the numerical studies are available in \url{https://github.com/cssr-tools/pofff}. The optimization tool Everest is available at \url{https://github.com/equinor/ert}. The Python image analysis toolbox DarSIA used for obtaining the segmentation and continuous CO$_2$ data can be accessed via \url{https://github.com/pmgbergen/DarSIA}; corresponding runscripts analyzing the FluidFlower dataset are available at \url{https://github.com/pmgbergen/fluidflower_benchmark_analysis}. The plotting tool for the simulation results is available at \url{https://github.com/cssr-tools/plopm}.

\paragraph{Funding} The authors acknowledge funding from Centre of Sustainable Subsurface Resources (CSSR), grant nr. 331841, supported by the Research Council of Norway, research partners NORCE Research AS and the University of Bergen, and user partners Equinor ASA, Harbour Energy, Sumitomo Corporation, Earth Science Analytics, GCE Ocean Technology, and SLB Scandinavia. JWB was in part supported by the Research Council of Norway through the project Unlocking maximal geological CO$_2$ storage through experimentally validated mathematical modeling of dissolution and convective mixing (project nr. 355188).

\begin{appendices}
\section{Configuration file format}\label{secA1}
Configuration files ease the setting and reproducibility of simulation studies. In \texttt{pofff}, the widely-used TOML format is adopted, see Figure \ref{confifile} for the initial configuration file run in this study.

\begin{figure}[h!]
\centering
\begin{lstlisting}[language=make,basicstyle=\miniscule]
# Set the full path to the flow executable and flags
flow="flow --enable-opm-rst-file=true --enable-tuning=true --linear-solver=cpr_trueimpes"

# Set the model parameters
grid="corner-point" # Type of grid (cartesian, tensor, or corner-point (cp))
thickness="final" # Thickness maps (measured 'initial', 'final', or a real positive value)
mult_thickness=1 # Thickness multiplier (a real positive value)
x=[140] # If cartesian, number of x cells; otherwise, variable array of x-refinement
# cartesian, number of z cells; tensor, variable array of refinement; cp, fix array of refinement (18 entries)
z=[7,5,5,5,5,5,5,8,10,9,5]
temperature=[20, 20] # Temperature bottom and top rig [C]
pressure=104900 # Pressure at the datum [Pa]           
diffusion=[1e-9, 2e-8] # Diffusion (in liquid and gas) [m^2/s]
sources=[[0.9, 0.005, 0.3], [1.7, 0.005, 0.7]] # Source positions: x, y, and z coordinates [m], source 1 to 2

# Set the saturation functions
krw="(max(0, (sw - swi) / (1 - swi))) ** nkrw"      # Wetting rel perm saturation function [-]
krn="(max(0, (1 - sw - sni) / (1 - sni))) ** nkrn"  # Non-wetting rel perm saturation function [-]
cap="pen * ((sw-swi) / (1-swi)) ** (-(1.0 / npen))" # Capillary pressure saturation function [Pa]

# Facie properties (ESF, C, D, E, F, G -> facie1, facie2, ..., facie6)
facie1={"PERMX1"=50E3,"PERMZ1"=50E3,"PORO1"=0.37,"DISPERC1"=0,"SWI1"=0.32,"SNI1"=0.1,"PEN1"=1500,"NKRW1"=2,"NKRN1"=2,"NPE1"=2,"THRE1"=5e-2,"NPNT1"=100}
facie2={"PERMX2"=100E3,"PERMZ2"=100E3,"PORO2"=0.38,"DISPERC2"=0,"SWI2"=0.14,"SNI2"=0.1,"PEN2"=800,"NKRW2"=2,"NKRN2"=2,"NPE2"=2,"THRE2"=5e-2,"NPNT2"=100}
facie3={"PERMX3"=300E3,"PERMZ3"=300E3,"PORO3"=0.40,"DISPERC3"=0,"SWI3"=0.12,"SNI3"=0.1,"PEN3"=200,"NKRW3"=2,"NKRN3"=2,"NPE3"=2,"THRE3"=5e-2,"NPNT3"=100}
facie4={"PERMX4"=800E3,"PERMZ4"=800E3,"PORO4"=0.39,"DISPERC4"=0,"SWI4"=0.12,"SNI4"=0.1,"PEN4"=150,"NKRW4"=2,"NKRN4"=2,"NPE4"=2,"THRE4"=5e-2,"NPNT4"=100}
facie5={"PERMX5"=1500E3,"PERMZ5"=1500E3,"PORO5"=0.39,"DISPERC5"=0,"SWI5"=0.12,"SNI5"=0.1,"PEN5"=100,"NKRW5"=2,"NKRN5"=2,"NPE5"=2,"THRE5"=5e-2,"NPNT5"=100}
facie6={"PERMX6"=3000E3,"PERMZ6"=3000E3,"PORO6"=0.42,"DISPERC6"=0,"SWI6"=0.1,"SNI6"=0.1,"PEN6"=1,"NKRW6"=2,"NKRN6"=2,"NPE6"=2,"THRE6"=5e-2,"NPNT6"=100}

# Schedule: 1) injection time [s], 2) time step size to write results [s], 3) injection rate [kg/s] (source1),
# 4) injection rate [kg/s] (source2). If --enable-tuning=true, then the TUNING values [days]
inj=[[  8100,  8100, 3E-7,    0, '1e-2 3e-4 1e-20 1e-20 1.6 0.2 0.65 1.1'], 
     [ 10200, 10200, 3E-7, 3E-7, '1e-2 1e-4 1e-20 1e-20 1.6 0.2 0.65 1.1'], 
     [ 68100, 68100,    0,    0, '1e-2 1e-3 1e-20 1e-20 1.6 0.2 0.65 1.1'],
     [345600, 86400,    0,    0, '1e-2 1e-2 1e-20 1e-20 1.6 0.2 0.65 1.1']]

# Everest
min_realizations_success = 0 # Minimum number of simulations to be regarded as a success.
max_function_evaluations = 200000 # Maximum number of simulations
random_seed = 7 # Set a specific seed for reproducibility; a value of 0 means no seed
cores = 50 # Maximum number of simulations running in parallel
maxtime = 3600 # Maximum runtime in seconds of a realization; a value of 0 means unlimited runtime
delete = true # Delete large files?
monotonic = false # Only consider monotonic values, e.g, increasing entry pressure with decreasing sand size
popsize = 200 # Population size
PERM2 = [ 100E3,  50E3,  200E3, 50] # Initial value, min, max, and size of interval
PERM3 = [ 300E3, 100E3,  500E3, 50]
PERM4 = [ 800E3, 300E3, 1300E3, 50]
PERM5 = [1500E3, 800E3, 3000E3, 50]
SWI2 =  [  0.14,  0.05,   0.35, 10]
SWI3 =  [  0.12,  0.05,   0.32, 10]
SWI4 =  [  0.12,  0.05,   0.30, 10]
SWI5 =  [  0.12,  0.05,   0.28, 10]
SNI2 =  [  0.10,  0.05,   0.35, 10]
SNI3 =  [  0.10,  0.05,   0.32, 10]
SNI4 =  [  0.10,  0.05,   0.30, 10]
SNI5 =  [  0.10,  0.05,   0.28, 10]
PEN2 =  [   800,   500,   1000, 20]
PEN3 =  [   200,   150,    600, 20]
PEN4 =  [   150,   100,    400, 20]
PEN5 =  [   100,    50,    200, 20]
\end{lstlisting}
\caption{Example of configuration file for \texttt{pofff}. This file can be run from the terminal as: pofff -i input\_file.toml -o output\_folder -m everest -t 24,48,72,96,120, where -t sets the times in days for the experimental images to HM}
\label{confifile} 
\end{figure}

In Figure \ref{confifile} comments are added to describe the different available settings, and we refer the reader to the \texttt{pofff} online documentation for an extended description. We remark that this approach (using configuration files to set up numerical studies) easily allows for further extension of the work.

\section{Lower threshold for CO$_2$ concentration}\label{secA2}
\begin{figure}[h!]
\centering
\includegraphics[width=\textwidth]{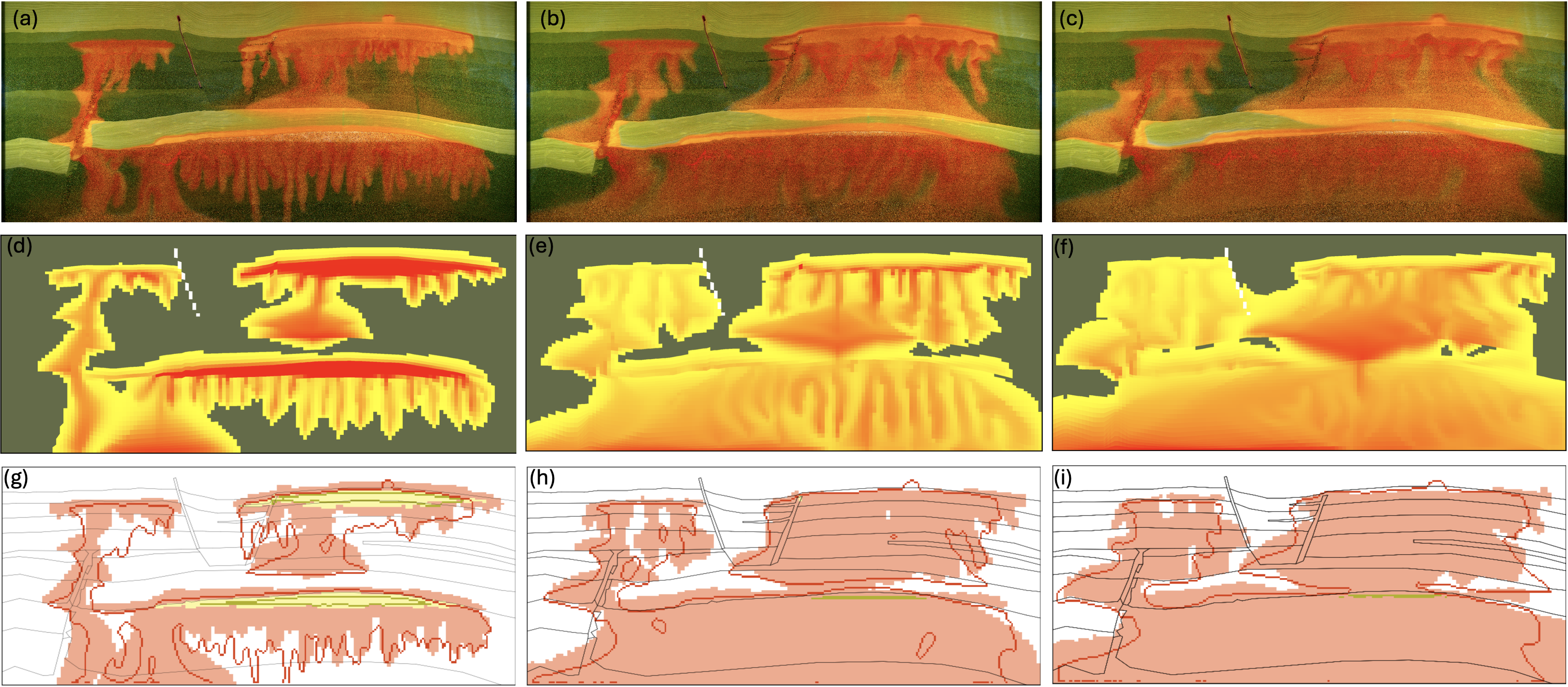}
\caption{(a-c) Photographs from the FluidFlower experiment C2 at one, three, and five days, respectively. (d-e) Spatial maps of CO$_2$ concentration from the corresponding simulation results. (f-g) Contour plots comparing experimental observations and segmented simulation outputs (threshold of 0.05 kg/m$^3$ for the CO$_2$ concentration) at the same time intervals}
\label{aff_maps}
\end{figure}

\begin{figure}[h!]
\centering
\includegraphics[width=\textwidth]{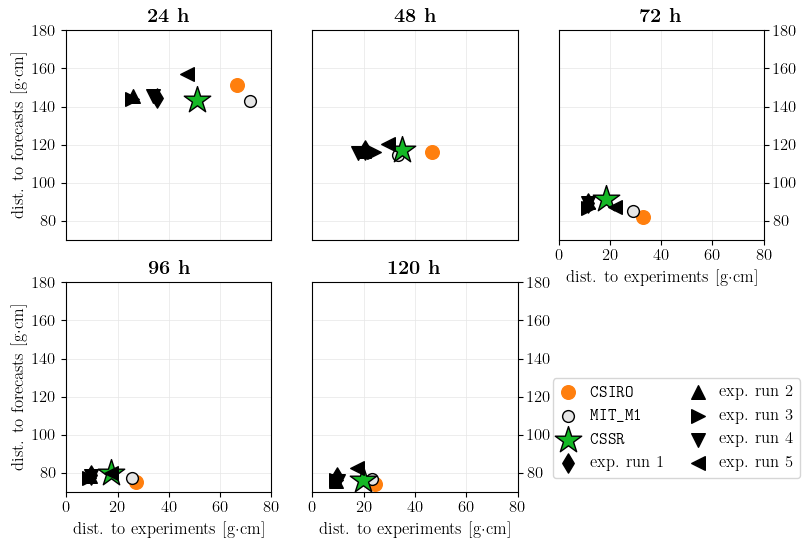}
\caption{Wasserstein distances computed between simulation results and experimental data (experiments C1-C5), as well as between simulation forecasts (threshold of 0.05 kg/m$^3$ for the CO$_2$ concentration)}
\label{aff_emd}
\end{figure}

To illustrate the sensitivity of the segmentation threshold in the numerical simulations of CO$_2$ concentration, Figures \ref{aff_maps} and \ref{aff_emd} present the spatial distribution maps and corresponding Wasserstein distances obtained using a threshold of 0.05 kg/m$^3$, in contrast to the benchmark value of 0.1 kg/m$^3$. See Table \ref{aerrors} for the computed errors for the sparse data and Wasserstein distances, and Table \ref{aparameters} for the simulation parameters used to generate these results.

\begin{table}[h]
\centering
\caption{Sparse data and Wasserstein distance (WD) comparison between experimental and simulation results (threshold of 0.05 kg/m$^3$ for the CO$_2$ concentration)}\label{aerrors}
\begin{tabular}{@{}lrrrrrrrrr@{}}		
\toprule
Data source & \# cells & 2 (s )& 3a (g) & 3c (g) & 3d (g) & 4c (g) & 6 (g) & error$_\text{g}$ & WD (g$\cdot$cm)\\
\midrule
Experiment & - & 14800 & 0.23 & 3.10 & 0.38 & 0.78 & 0.57 & - & 18.32 \\
CSIRO & 151,402 & 18000 & 0.35 & 3.60 & 0.57 & 0.30 & 0.57 & 33.38 & 39.52 \\
MIT\_M1 & 44,284 & 17700 & 0.87 & 2.11 & 0.43 & 0.52 & 0.52 & 63.60 & 36.31 \\
CSSR & 9,627 & 15600 & 0 & 3.40 & 0.31 & 0.44 & 0.50 & 31.61 & 28.32 \\
\bottomrule
\end{tabular}
\end{table}

\begin{table}[h]
\centering
\caption{Model parameters}\label{aparameters}
\begin{tabular}{@{}lrrrrrr@{}}		
\toprule
Id & Sand&$k$ [D]&$\phi$  [-] &$s_{r,w}$ [-]&  $s_{r,g}$ [-] & $p_e$ [Pa]\\
\midrule
1 & ESF	&62	 	&0.37&0.34& 0.25& 1900\\
2 & C	&152	&0.38&0.32& 0.20& 600\\
3 & D	&690	&0.40&0.30& 0.19& 305\\
4 & E	&720	&0.39&0.28& 0.06& 175\\
5 & F	&2000 	&0.39&0.25& 0.01& 170\\
6 & G	&2500	&0.42&0.17& 0& 55\\
\bottomrule
\end{tabular}
\end{table}

\section{Computational details}\label{secA3}
This appendix provides a detailed description of the numerical study setup. The configuration files, scripts, and supplementary technical information required to reproduce the results are available in the \texttt{pofff} repository and its online documentation. The procedures outlined here are based on the Ubuntu local server configuration employed in our work. For users operating on macOS or Windows (via the Windows Subsystem for Linux), supplementary instructions are available in the official \texttt{pofff} online documentation to facilitate replication of the simulations across different platforms.

The simulations were executed on an Intel\textregistered\; Xeon\textregistered\; Platinum 8354H processors, featuring a 3.1 GHz base clock, x86\_64 architecture, and a total of 144 logical CPUs (4 sockets $\times$ 18 cores per socket, with 2 threads per core). The system is equipped with 99 MB of L3 cache, 1583 GB RAM, and an HDD-based storage solution. It operated on Ubuntu 24.04.3 LTS (Noble), with Open MPI version 4.1.6 installed. Although GPU acceleration was not utilized in the simulations, it is worth noting that OPM Flow provides GPU support. Readers interested in leveraging this capability are encouraged to consult the OPM Flow manual (\cite{flowmanual}) for further details.

The numerical experiments were conducted using OPM Flow version 2025.10. The simulations employed the default Python distribution in Ubuntu 24.04 LTS, corresponding to Python 3.12.3. The \texttt{pofff} package was also used in version 2025.10. A complete specification of Python dependencies and their exact versions is provided in the pyproject.toml file included with the \texttt{pofff} v2025.10 release. For reproducibility, the \texttt{pofff} repository contains a continuous integration script (CI.yml) located in .github/workflows. This script is configured to run on the latest available Ubuntu release (at the time of writing, Ubuntu 24.04 LTS) with Python 3.12. Following the steps in this workflow allows users to install the OPM Flow binaries, create a dedicated Python virtual environment, install \texttt{pofff} together with the required Python libraries, and execute the test cases to verify that the installation has been successfully completed.

\begin{figure}[h!]
\centering
\includegraphics[width=0.7\textwidth]{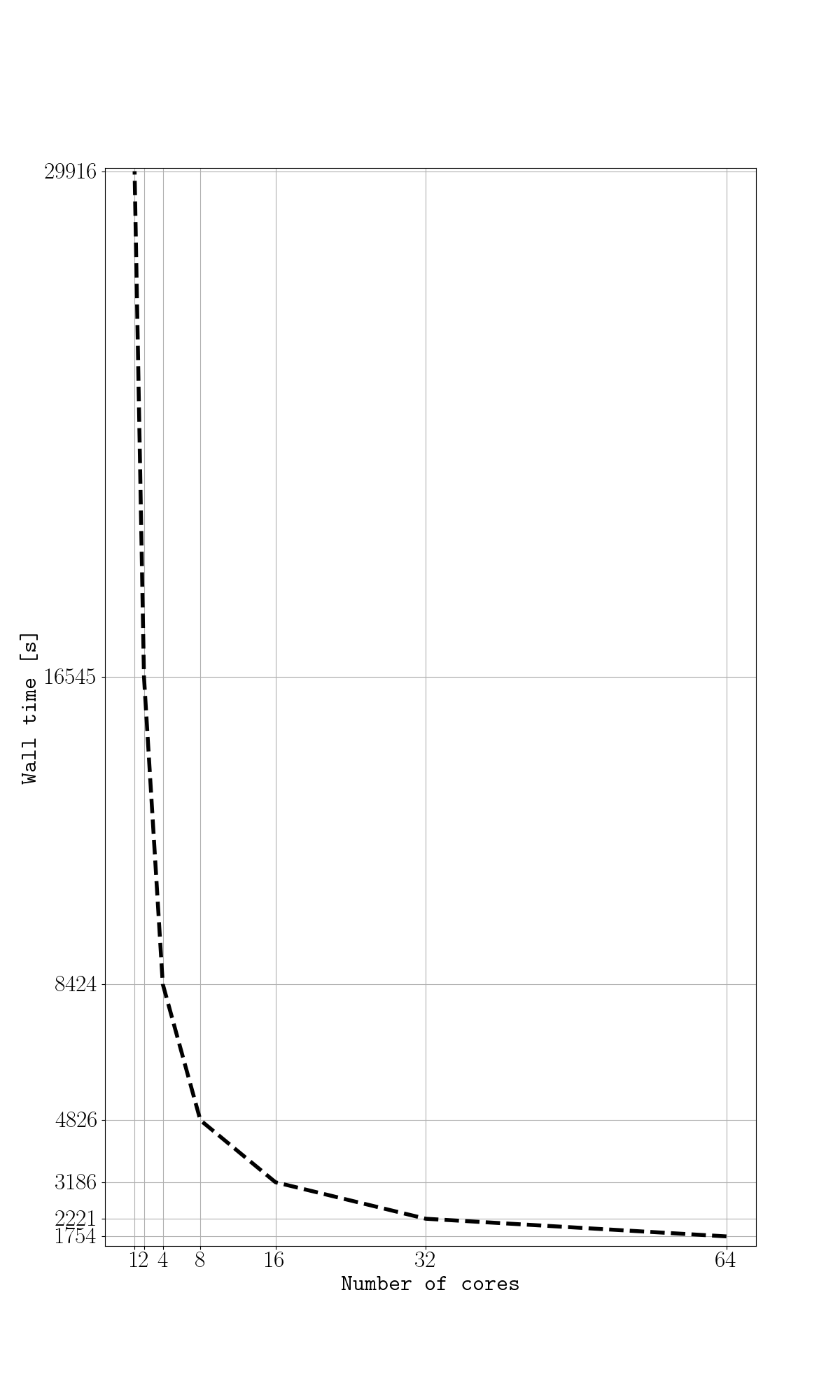}
\caption{Scaling curve of wall time as a function of number of cores}
\label{acpus}
\end{figure}

We profile the integrated workflow in three stages: preprocessing, history matching (HM), and postprocessing. The preprocessing stage consists of generating input files for both OPM Flow and the history matching tool. The HM stage involves executing a sequence of tasks, including OPM Flow simulations, for which profiling results are reported separately for the simulations and the remaining HM tasks. The postprocessing stage comprises the generation of figures and tables. For this profiling study, we use the configuration file from the final HM iteration presented in this paper, but restrict the run to 64 simulations to ensure computational feasibility. The case is executed on a single core, with reporting times summarized in Table \ref{aprofiling}, while scalability with respect to the number of cores is illustrated in Figure \ref{acpus}.

\begin{table}[h]
\caption{Execution time profiling of the integrated \texttt{pofff} workflow}\label{aprofiling}
\centering	
\begin{tabular}{@{}lccccc@{}}		
\toprule
 & Preprocessing & \multicolumn{2}{c}{History matching}& Postprocessing & Total\\
 &  & OPM Flow & Rest & & \\
\midrule
Wall time (s) & 8.35 & 27,491.92\footnotemark[1] & 1,928.21 & 487.14 & 29,915.63 \\
\bottomrule
\end{tabular}
\\$^1$ 432.99$\pm$7.60 (median and interquantile range of the 64 OPM Flow runs).\hspace{1.4cm}
\end{table}

\noindent Complementing the sensitivity studies reported in the FluidFlower benchmark special issue (\cite{Wang2024} on hysteresis and molecular diffusion, \cite{Wapperom2024} on discretization techniques and types of grids, \cite{Jammoul2024} on rock and gas relative permeability, and image segmentation threshold), as well as the OPM team's results for the SPE11 benchmark (\cite{nordbotten2025,Landa-Marbán2025}) on types of grids, grid resolution, and solver tolerances, we investigate sensitivities in the HM related to the random seed and the operating system. The base case employs a random seed of 7, executed on the Ubuntu machine described earlier. This configuration corresponds to the case presented in Table \ref{aprofiling} and Figure \ref{acpus}. To assess sensitivity, we vary one factor at a time: the random seed is changed to 11, and the operating system is switched to macOS (Tahoe 26.1, Apple M2 Pro chip). Table \ref{asensitivity} reports the Wasserstein distance for each case. These results highlight the sensitivity of the HM outcomes to choices of random seed and operating system.

\begin{table}[h]
\caption{Sensitivity of the history matching to setup choices}\label{asensitivity}
\centering	
\begin{tabular}{@{}lccc@{}}		
\toprule
& Base case & Random seed & Operating system \\
\midrule
Wasserstein distance [g$\cdot$cm] & 32.37 & 32.71 & 33.19 \\
\bottomrule
\end{tabular}
\end{table}

\noindent Regarding the accuracy-time trade-off analysis, we adopt the CO$_2$ mass in the system as the evaluation metric. Let CO$_2^{\text{inj}}$ denote the total injected CO$_2$ mass (8.5 grams), and CO$_2^{\text{sim}}$ the CO$_2$ mass at the end of the simulation. The accuracy is then defined as:
\begin{equation}
\text{Accuracy}=100\times\left(1-\frac{|\text{CO}_2^{\text{sim}}-\text{CO}_2^{\text{inj}}|}{\text{CO}_2^{\text{inj}}}\right).
\end{equation}
Table \ref{atradeoff} reports the accuracy metric evaluated under two convergence (CNV) tolerance settings, spanning from the relaxed default values in OPM Flow (-{}-tolerance-cnv=0.01 and -{}-tolerance-cnv-relaxed=1) to tightened tolerances (-{}-tolerance-cnv=0.001 and -{}-tolerance-cnv-relaxed=0.001).

\begin{table}[h]
\caption{Accuracy-time trade-off for CO$_2$ mass}\label{atradeoff}
\centering	
\begin{tabular}{@{}lcrr@{}}	
\toprule
Case & OPM Flow parameters &  Accuracy ($\%$) & Wall time (s)\\
\midrule
Relax\footnotemark[1] & -{}-tolerance-cnv=0.01 -{}-tolerance-cnv-relaxed=1 & 99.987	&417.52	\\
Tight & -{}-tolerance-cnv=0.001 -{}-tolerance-cnv-relaxed=0.001 & 99.998	&570.58 \\
\bottomrule
\end{tabular}
\\$^1$ Default values in OPM Flow.\hspace{9.8cm}
\end{table}

\noindent Table \ref{atradeoff} indicates that adopting the relaxed CNV tolerance provides a suitable balance for the simulations presented in this study. As described in the online documentation of \texttt{pyopmspe11} (\cite{Landa-Marbán2025}) for the SPE11a case with a 1 mm grid, using the default relax tolerances yields a mass accuracy of 96.5\% with a simulation time of 17 days. By contrast, enforcing tighter tolerances improves the accuracy to 99.8\%, but at the expense of a substantially longer runtime of 55 days.

\end{appendices}

\bibliography{bibliography}

\end{document}